\documentclass[aps,prb,amsfonts,amssymb,twocolumn,amsmath,preprintnumbers,nofootinbib,floatfix,showpacs,superscriptaddress]{revtex4-1} 
\usepackage[pdftex]{graphicx}
\usepackage[utf8]{inputenc}
\usepackage{bbold}
\usepackage{dcolumn}
\usepackage{bm}
\usepackage{epsfig}
\usepackage{latexsym} 
\usepackage{amsmath}
\usepackage{amsfonts}
\usepackage{amssymb}
\usepackage{xcolor}
\usepackage{subfigure}
\usepackage{array}
\usepackage{bbm}
\usepackage{hyperref,hypcap}
\usepackage{cancel}
\usepackage{ulem}
\usepackage{graphicx}
\graphicspath{{images/}}
\usepackage{braket}
\usepackage{commath}
\providecommand{\U}[1]{\protect\rule{.1in}{.1in}}

\usepackage{booktabs}
\newcolumntype{C}[1]{>{\centering\arraybackslash}m{#1}}
\AtBeginDocument{
\heavyrulewidth=.08em
\lightrulewidth=.05em
\cmidrulewidth=.03em
\belowrulesep=.65ex
\belowbottomsep=0pt
\aboverulesep=.4ex
\abovetopsep=0pt
\cmidrulesep=\doublerulesep
\cmidrulekern=.5em
\defaultaddspace=.5em
}
\newcolumntype{R}[1]{>{\raggedleft\arraybackslash}p{#1}}

\newcommand{\<}{\langle}

\renewcommand{\>}{\rangle}
\renewcommand{\(}{\left(}
\renewcommand{\)}{\right)}
\renewcommand{\[}{\left[}
\renewcommand{\]}{\right]}

\renewcommand{\d}{\partial}

\newcommand{\angstrom}{\textup{\AA}}

\begin{document}

\title{Lattice dynamics with molecular Berry curvature: chiral optical phonons}

\author{Daniyar Saparov} 
\affiliation{Department of Physics, The University of Texas at Austin, Austin, Texas 78712, USA}

\author{Bangguo Xiong} 
\affiliation{Department of Physics, The University of Texas at Austin, Austin, Texas 78712, USA}

\author{Yafei Ren}
\email{yfren@uw.edu}
\affiliation{Department of Physics, The University of Texas at Austin, Austin, Texas 78712, USA}
\affiliation{Department of Materials Science and Engineering, University of Washington, Seattle, Washington 98195, USA}

\author{Qian Niu} 
\affiliation{Department of Physics, The University of Texas at Austin, Austin, Texas 78712, USA}
\affiliation{ICQD/HFNL and School of Physics, University of Science and Technology of China, Hefei, Anhui 230026, China}

\date{\today}

\begin{abstract}
Under the Born-Oppenheimer approximation, the electronic ground state evolves adiabatically and can accumulate geometrical phases characterized by the molecular Berry curvature. In this work, we study the effect of the molecular Berry curvature on the lattice dynamics in a system with broken time-reversal symmetry. The molecular Berry curvature is formulated based on the single-particle electronic Bloch states. It manifests as a \textit{non-local} effective magnetic field in the equations of motion of the ions that are beyond the widely adopted Raman spin-lattice coupling model. We employ the Bogoliubov transformation to solve the quantized equations of motion and to obtain phonon polarization vectors. We apply our formula to the Haldane model on a honeycomb lattice and find a large molecular Berry curvature around the Brillouin zone center. As a result, the degeneracy of the optical branches at this point is lifted intrinsically. The lifted optical phonons show circular polarizations, possess large phonon Berry curvature, and have a nearly quantized angular momentum that modifies the Einstein-de Haas effect.
\end{abstract}

\maketitle

\section{Introduction}
The Born-Oppenheimer approximation assumes an adiabatic evolution of electronic states following motion of the ions~\cite{HuangBook}. During the evolution, the electronic ground state can accumulate nontrivial geometrical phase in the absence of time-reversal symmetry\cite{MeadTruhlar, MeadTruhlarRMP}. The influence of this phase on the ion's dynamics was discussed first by Mead and Truhlar in molecules\cite{MeadTruhlar}, which was later identified as an electronic Berry phase with respect to the ion's displacement~\cite{MeadTruhlarRMP} and dubbed as a molecular Berry phase\cite{NameMBC}. In magnetic molecules, the molecular Berry phase can induce vibration modes with nonzero angular momentum~\cite{ChiralMolecule}. 

In a periodic lattice, the molecular Berry curvature associated with this phase can influence directly the lattice dynamics and therefore the properties of the phonons~\cite{TQin2012, Qin2011, Qi11, PhononVisco_16, PhononHallVisco_Optical_19, PhononMagnetoChiral_20, PhononChiralAnomaly_17, PhononHelicity_Electronic_21}. In the long-wavelength limit, this Berry curvature manifests as a Hall viscosity~\cite{Qi11, PhononVisco_16, PhononHallVisco_Optical_19} that can modify the dispersion, polarization, and transport properties of the long-wavelength phonons~\cite{TQin2012, Qin2011, PhononMagnetoChiral_20, PhononChiralAnomaly_17}.  
By considering a finite overlap between electronic wavefunctions on neighboring sites, a recent work studied the molecular Berry curvature induced by a magnetic field $B$ in a nonmagnetic insulator in the linear order of $B$~\cite{Saito2019}. However, the molecular Berry curvature in a Bloch system without a uniform magnetic field has not been explicitly studied~\cite{Price2014}. A Bloch-wavefunction-based formula of the molecular Berry curvature is highly desired~\cite{DFTthermalHall}. 

In this work, we explore the effect of molecular Berry curvature on the lattice dynamics in the absence of a uniform magnetic field. In an electronic system that breaks the time-reversal symmetry and respects the translational symmetry we formulate the molecular Berry curvature by using single-particle Bloch wavefunctions and assuming the many-body electronic ground state as a Slater determinant. The molecular Berry curvature influences the lattice dynamics as an effective magnetic field, which however is nonlocal. We then employ the Bogoliubov transformation to solve the quantized equation of motions and to obtain the spectrum and polarization vector of the phonons. 

We apply the formula to the Haldane model of a honeycomb lattice. The molecular Berry curvature exhibits a peak value at the Brillouin zone center. The peak value depends strongly on the electronic band gap and the electronic band topology. The narrow distribution of the molecular Berry curvature in momentum space indicates that, in real space, one atom can be influenced by the velocity of another atom far away. With the molecular Berry curvature, the double degeneracy of the optical phonons at the Brillouin zone center is lifted intrinsically, in contrast to the splitting induced extrinsically by the magnetic field~\cite{PhononMagnetization_20, PhononMagPbTe, Phonon_Spin_20, Phonon_OrbitMoment_19, Cong_20, Dong_Niu_18, OrbMag_Adiabatic_19, PhononMag}. The polarization vectors become left and right-handed, separately, which carry nonzero angular momenta contributing to a nonzero zero-point angular momentum of the lattice vibration~\cite{LZhang2014}. 
The phonon modes also carry nonzero phonon Berry curvature and contribute to the phonon thermal Hall effect, which attracts much attention in experiments recently~\cite{Strohm2005, Inyushkin2007, PHE_SpinLiquid_Exp_17, ChiralPhononCuprate_NP_20}.  

\section{MOLECULAR BERRY CURVATURE AND PHONON POLARIZATION}
\subsection{Molecular Berry curvature}
Under the Born-Oppenheimer approximation, electrons stay at their instantaneous ground state $|\Phi_0(\{\bm{R}\})\>$ at a given time with lattice configuration $\{\bm{R}\}$. 
When the lattice configuration evolves, the electronic ground state evolves adiabatically and accumulates a geometrical phase, which in turn can modify the lattice dynamics. The geometrical phase manifests itself as a gauge field $\bm{A}_{l, \kappa}$ in the lattice Hamiltonian (it was originally reported by Mead and Truhlar~\cite{MeadTruhlar} and an alternative derivation is shown in Appendix A)
\begin{align} \label{eq1}
H_L = \sum_{l, \kappa} \frac{1}{2 M_{\kappa}} \(\bm{p}_{l, \kappa} - \hbar \bm{A}_{l, \kappa}(\{\bm{R}\})\)^2 + V_{\text{eff}} (\{\bm{R}\})
\end{align}
where $\bm{p}_{l,\kappa} = - i \hbar \bm{\nabla}_{l, \kappa}$ with $\bm{\nabla}_{l, \kappa} = \d/\d \bm{R}_{l, \kappa}$ is the canonical momentum of the $\kappa$-th atom at the $l$-th unit cell with a coordinate $\bm{R}_{l, \kappa}$ and a mass $M_{\kappa}$. The scalar potential $V_{\rm eff}(\{\bm{R}\})$ is contributed from the Coulomb interaction of the ions and the electrons whereas the vector potential $\bm{A}_{l, \kappa}(\{\bm{R}\}) = i \<\Phi_0(\{\bm{R}\})|\bm{\nabla}_{l, \kappa} \Phi_0(\{\bm{R}\}) \>$ is the molecular Berry connection that describes the geometrical phase of the electronic ground state. Although the molecular Berry connection is gauge dependent, it can give rise to a gauge invariant molecular Berry curvature \cite{TQin2012, DXiao2010}
\begin{gather} \label{eq2}
G^{\kappa \alpha}_{\kappa' \beta} (\bm{R}_{l},\bm{R}_{l'})=2 \text{Im}\Big\<\frac{\d \Phi_0}{\d R_{l', \kappa'\beta}} \Big| \frac {\d \Phi_0}{\d R_{l, \kappa\alpha}} \Big \> 
\end{gather}
where the indices $\alpha, \beta$ represent the Cartesian components of the coordinates. One can proceed further under the assumption that every lattice point vibrates around its equilibrium position with $\{\bm{R}\} = \{\bm{R}^0_{l,\kappa} + \bm{u}_{l, \kappa}, l = 1, \dots, N; \kappa = 1, \dots, r \}$ where the equilibrium position $\bm{R}^0_{l, \kappa} \equiv \bm{R}^0_{l} + \bm{d}_{\kappa}$ with $\bm{R}^0_l$ being the equilibrium position of the $l$-th unit cell, $\bm{d}_{\kappa}$ being the relative position of the $\kappa$-th ion, and $\bm{u}_{l,\kappa}$ being its displacement. At the equilibrium configuration, the Berry curvature $G^{\kappa \alpha}_{\kappa' \beta} (\bm{R}_{l},\bm{R}_{l'})$ exhibits translational symmetry that depends on $\bm{R}^0_{l}-\bm{R}^0_{l'}$ only. 

In the following, we consider a symmetric gauge~\cite{Holz1972}, which exists near the equilibrium position as shown in Appendix B, such that
\begin{gather} \label{eq3}
A_{l,\kappa\alpha} = -\frac{1}{2} \sum_{\kappa', \beta, l'} G^{\kappa \alpha}_{\kappa' \beta} (\bm{R}^0_{l}-\bm{R}^0_{l'})  u_{l',\kappa'\beta}.
\end{gather}
By taking the advantage of the translational invariance, we express the lattice Hamiltonian in momentum space
\begin{align} \nonumber
H_L = & \sum_{\bm{k}, \kappa} \frac{1}{2 M_{\kappa}} \[ \bm{p}_{\kappa}(-\bm{k}) - \hbar \bm{A}_{\kappa}(-\bm{k})\] \[ \bm{p}_{\kappa}(\bm{k}) - \hbar \bm{A}_{\kappa}(\bm{k})\] \\ \label{eq4}
& + V_{\text{eff}}\(\{\bm{u}(\bm{k})\}\)
\end{align}
where the momentum-space Berry connection
\begin{align*}
 {A}_{\kappa\alpha}(\bm{k})\doteq \frac{1}{\sqrt{N}}\sum_{l} {A}_{l, \kappa\alpha} e^{-i\bm{k}\cdot \bm{R}^0_{l}} = -\frac 12 \sum_{\kappa', \beta} G^{\kappa \alpha}_{\kappa' \beta} (\bm{k}) u_{\kappa'\beta}(\bm{k})   \nonumber
\end{align*}
with
\begin{equation} \label{eq5}
\begin{aligned} 
  \bm{u}_{\kappa}({\bm{k}}) &= \frac{1}{\sqrt{N}} \sum_{l} \bm{u}_{l, \kappa} e^{-i\bm{k}\cdot \bm{R}^0_{l}} \\
  \bm{p}_{\kappa}({\bm{k}}) &= \frac{1}{\sqrt{N}} \sum_{l} \bm{p}_{l, \kappa} e^{-i\bm{k}\cdot \bm{R}^0_{l}} \\
  G^{\kappa \alpha}_{\kappa' \beta} (\bm{k}) &= \frac{1}{N}\sum_{l} \sum_{l'} G^{\kappa \alpha}_{\kappa' \beta} (\bm{R}^0_{l}-\bm{R}^0_{l'}) e^{-i\bm{k}\cdot(\bm{R}^0_{l}-\bm{R}^0_{l'})}.
\end{aligned}
\end{equation}

We further express the momentum-space molecular Berry curvature in a gauge invariant form by employing a set of many-body wavefunction $\{|\Phi_n\>\}$ with the completeness relation $\sum_{n} |\Phi_n\>\<\Phi_n| = 1$ and associated eigenenergy $E_n$. By using the identity $\<\Phi_n|\mathcal{M}_{\bm{k},\kappa\alpha}|\Phi_{n'}\> = \<\frac{\partial \Phi_n}{\partial u_{-\bm{k},\kappa\alpha}}|\Phi_{n'}\>(E_n - E_{n'}) $ for $ n' \neq n $, the molecular Berry curvature reads 
\begin{align} \nonumber
G^{\kappa \alpha}_{\kappa' \beta} (\bm{k}) &= i \sum_{n\neq0} \[\frac{\<\Phi_0 |\mathcal{M}_{\bm{k},\kappa\alpha} |\Phi_n\>\<\Phi_n |\mathcal{M}_{-\bm{k},\kappa'\beta} |\Phi_0\>}{(E_n-E_0)^2}\] \\
& - \{\mathcal{M}_{\bm{k},\kappa\alpha} \leftrightarrow \mathcal{M}_{-\bm{k},\kappa'\beta}\} \label{eq6}
\end{align} 
where $E_0$ is the energy of the electronic ground state, $E_n$ is for the excited states, $\mathcal{M}_{\bm{k},\kappa\alpha}=\frac{\partial H_{e}}{\partial u_{-\bm{k},\kappa\alpha}}|_{u_{-\bm{k},\kappa\alpha}\rightarrow 0}$ represents the electron-phonon coupling with $H_{e}$ being the electronic Hamiltonian that depends on the atomic coordinates. By further taking $\{|\Phi_n\>\}$ as Slater determinant, the above formula can be expressed in terms of single-particle Bloch wavefunctions as detailed in Appendix D. This expression can readily be applied to a specific model using the first-principles approach.

\subsection{Phonon Polarization Vectors}
We further simplify the notation by normalizing the coordinates and expressing the Hamiltonian in terms of matrices. We first define column vectors $p_{\bm{k}} = \( \dots p_{\kappa\alpha}(\bm{k})/\sqrt{M_\kappa} \dots \)^T$ and $u_{\bm{k}} = \( \dots \sqrt{M_\kappa} u_{\kappa\alpha}(\bm{k}) \dots \)^T$. For the two dimensional system studied in this work, there are $2r$ elements. We also define the matrix $\tilde{G}_{\bm{k}}$ with elements $\tilde{G}_{\bm{k}}(\kappa\alpha,  \kappa'\beta) = \frac{\hbar}{2\sqrt{M_{\kappa}M_{\kappa'}}} G^{\kappa \alpha}_{\kappa' \beta} (\bm{k})$ (see Appendix E). 
By expressing the potential energy in a quadratic form~\cite{Callaway1991}  $V_{\text{eff}}\(\{\bm{u}(\bm{k})\}\) = \frac12 u_{\bm{k}}^{\dagger} K_{\bm{k}} u_{\bm{k}}$, the lattice Hamiltonian reads
\begin{align} \label{eq7}
H_L & = \sum_{\bm{k}} \frac 12 \( \begin{matrix} u_{\bm{k}} \\ p_{\bm{k}} \end{matrix} \)^{\dagger} \( \begin{matrix} D_{\bm{k}} & \tilde{G}^{\dagger}_{\bm{k}} \\ \tilde{G}_{\bm{k}}  & \mathbb{1} \end{matrix} \) \( \begin{matrix} u_{\bm{k}} \\ p_{\bm{k}} \end{matrix} \)
\end{align}
where $D_{\bm{k}} = K_{\bm{k}} + \tilde{G}^{\dagger}_{\bm{k}} \tilde{G}_{\bm{k}}$. It is noted that $D_{\bm{k}}$ here is different from that in Ref.~\onlinecite{TQin2012}. The corresponding canonical equations of motion are \cite{Kittel1987}:
\begin{align} \label{eq8}
\( \begin{matrix}  
\dot{u}_{\bm{k}} \\
\dot{p}_{\bm{k}}
\end{matrix} \) = \( \begin{matrix}
\frac{\d H_L}{\d {p}_{-\bm{k}}} \\
- \frac{\d H_L}{\d u_{-\bm{k}}}
\end{matrix} \) = \( \begin{matrix}
\tilde{G}_{\bm{k}} & \mathbb{1} \\
- D_{\bm{k}} & \tilde{G}_{\bm{k}}
\end{matrix} \) \( \begin{matrix}
u_{\bm{k}} \\
p_{\bm{k}}
\end{matrix} \).
\end{align}

We then introduce the canonical transformation
\begin{align} \label{eq9}
u_{\bm{k}} &= \sum_{\nu} \sqrt{\frac{\hbar}{\omega_0}} \( \gamma^*_{\nu} b^{\dagger}_{-\bm{k}, \nu} + \gamma_{\nu} b_{\bm{k}, \nu} \) \\  \label{eq10}
p_{\bm{k}} &= \sum_{\nu} i \sqrt{\hbar \omega_0} \( \bar{\gamma}^*_{\nu} b^{\dagger}_{-\bm{k}, \nu} - \bar{\gamma}_{\nu} b_{\bm{k}, \nu}\)
\end{align} 
to diagonalize the Hamiltonian as 
\begin{align} \label{eq11}
 H_L = \sum_{\bm{k}, \nu} \hbar \omega_{\bm{k},\nu} \( b^{\dagger}_{\bm{k}, \nu} b_{\bm{k}, \nu} + \frac 12 \)
\end{align}
where $b^{\dagger}_{-\bm{k}, \nu}$ and $b_{\bm{k}, \nu}$ are the creation and annihilation operators of the phonon modes. These operators are constrained by the commutation relation 
$[b_{\bm{k}, \nu}, b^\dagger_{\bm{k}', \nu'}] = \delta_{\bm{k},\bm{k}'} \delta_{\nu,\nu'}$ and the Heisenberg equations of motion 
\begin{equation} \label{eq12} 
\begin{aligned} 
 \dot{b}_{\bm{k}, \nu} &= -i\omega_{\bm{k}, \nu} b_{\bm{k}, \nu} \\   
 \dot{b}^{\dagger}_{-\bm{k}, \nu} &= i\omega_{-\bm{k}, \nu} b^{\dagger}_{-\bm{k}, \nu}.
\end{aligned}
\end{equation}
Assisted by these identities, the phonon energy $\omega_{\bm{k},\nu}$ and the polarization vector $ \psi_\nu=\(\begin{matrix} \gamma_{\nu}, \bar{\gamma}_{\nu} \end{matrix}\)^T$ need to satisfy the eigenvalue equation 
\begin{align} \label{eq13}
 \omega_{\bm{k}, \nu} \psi_\nu = 
 \(\begin{matrix} i \tilde{G}_{\bm{k}} & \omega_0 \\ \frac{D_{\bm{k}}}{\omega_0} & i \tilde{G}_{\bm{k}} \end{matrix}\)
 \psi_\nu   
\end{align}
which can be obtained by substituting Eqs.~\eqref{eq9} and \eqref{eq10} into Eq.~\eqref{eq8}. The eigenvalues show particle-hole symmetry like property~\cite{Qin2011}, i.e.,
$\omega_{\nu,\bm{k}}=-\omega_{-\nu,-\bm{k}}$. Only the positive branches are physically allowed since only the wavefunctions for those branches can make the commutation relation $[b_{\bm{k}, \nu}, b^\dagger_{\bm{k}', \nu'}]=\delta_{\bm{k},\bm{k}'}\delta_{\nu,\nu'}$ valid with the normalization condition $\psi_\nu^\dagger \sigma_x \psi_\nu=1$.

One can transform the non-Hermitian problem to a Hermitian one. By multiplying Eq. (\ref{eq13}) with $\sigma_x$ from left, one can find that
\begin{align} \label{eq14}
\omega_{\bm{k}, \nu} \sigma_x \psi_{\nu} = \Omega_{\bm{k}} \psi_{\nu} 
\end{align}
with the Hermitian matrix
\begin{align}
\Omega_{\bm{k}} =\(\begin{matrix} \frac{D_{\bm{k}}}{\omega_0} & -i \tilde{G}_{\bm{k}}^\dagger \\ i \tilde{G}_{\bm{k}} & \omega_0 \end{matrix}\). \nonumber 
\end{align}
Multiplying Eq.~\eqref{eq14} with $\Omega_{\bm{k}}^{\frac{1}{2}}\sigma_x$ from left side and introducing a new set of eigenstates $\tilde{\psi}_{\nu} = \Omega_{\bm{k}}^{\frac{1}{2}} \psi_{\nu}$, where $\Omega_{\bm{k}}^{\frac{1}{2}}$ is also Hermitian, we come to a Hermitian eigenvalue problem as
\begin{align} \label{eq15}
\omega_{\bm{k}, \nu} \tilde{\psi}_{\nu} = \Omega_{\bm{k}}^{1/2} \sigma_x \Omega_{\bm{k}}^{1/2} \tilde{\psi}_{\nu} = H_{\text{eff}} \tilde{\psi}_{\nu}
\end{align}
where the effective Hamiltonian ${H}_{\text{eff}}$ is Hermitian.  

\section{Lattice dynamics in HALDANE MODEL}
\subsection{Electronic Model and Molecular Berry curvature}
In this section, we present a case study on the dynamics of the honeycomb lattice. In the harmonic approximation, the atoms are considered connected by springs with longitudinal and transverse spring constants $K_L$ and $K_T$, respectively. Details of this model can be found in Appendix E.

\begin{figure} [h]
\includegraphics[scale=0.28]{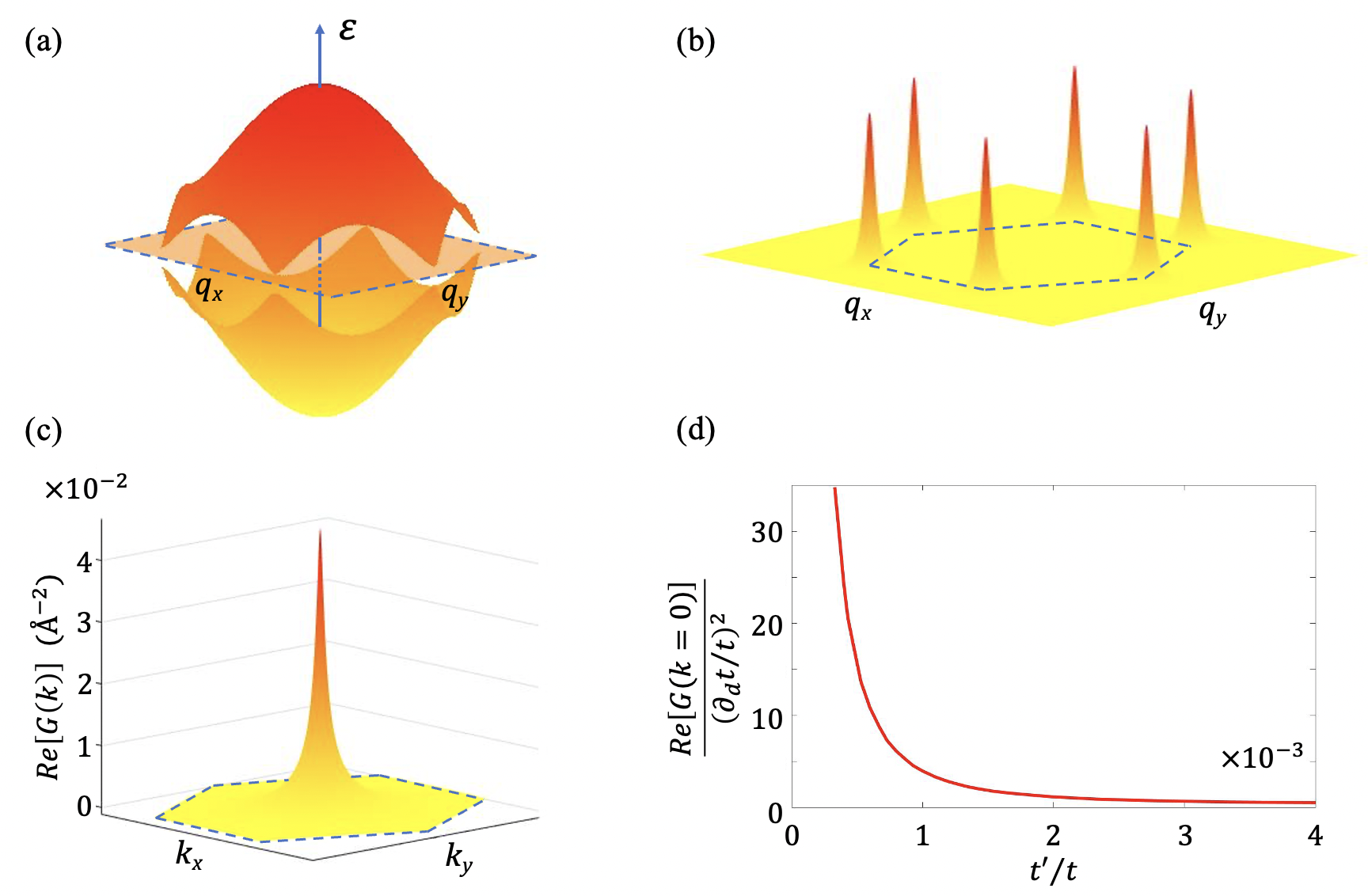}
\caption{(a) Electronic band structure represented in the Brillouin zone. Gap openings at the $K$ and $K'$ points are due to broken time-reversal symmetry. (b) Real part of the electronic contribution to the molecular Berry curvature $G^{Ax}_{Ay}$ at the $k=0$ limit ($\Gamma$ point). (c) Real part of the molecular Berry curvature $G^{Ax}_{Ay}$ in the phonon Brillouin zone. Largest electronic contributions come from the $K$ and $K'$ points (see FIG. 1b). (d) Dependence of the peak of the Berry curvature $G^{Ax}_{Ay}(\bm{k}=0)$ on the lattice parameters $\partial_d t,  t$, and $t'$.}
\label{FigBC}
\end{figure}

The time-reversal symmetry is broken by the electronic property described by the Haldane model~\cite{Haldane1988} with a tight-binding Hamiltonian
\begin{equation} \label{eq16}
\begin{aligned} 
H_{e}  = & - \sum_{\<i,j\>} t  a_i^{\dagger} b_j + \text{h.c.} \\ 
& - \sum_{\<\<i,j\>\>} t' e^{i \phi_{ij}} a_i^{\dagger} a_j 
 - \sum_{\<\<i,j\>\>} t' e^{-i \phi_{ij}} b_i^{\dagger} b_j \\ 
 = &\sum_{\bm{q}} \( \begin{matrix}
 a^{\dagger}_{\bm{q}} & b^{\dagger}_{\bm{q}}
 \end{matrix} \) \mathcal{H} (\bm{q})  \( \begin{matrix}
a_{\bm{q}} \\
b_{\bm{q}} 
\end{matrix} \)
\end{aligned}
\end{equation}
where $a_i$ ($a_i^{\dagger}$) and $b_i$ ($b_i^{\dagger}$) are electron creation (annihilation) operators of $A$ and $B$ sublattices respectively in the $i$-th unit cell. The first line represents the nearest neighbor hopping with the hopping energy $t$ while the second line represents the next-nearest neighbor hopping with a flux $\phi_{ij}$ attached to it. We set $\phi_{ij} = \pm \pi/2$ for clockwise/anti-clockwise hoppings. The lattice Hamiltonian can also be expressed in momentum space with kernel $\mathcal{H} (\bm{q})$ and $\bm{q}$ running over the first Brillouin zone. The single-particle Bloch eigenstates for the conduction and valence bands are denoted as $\phi^{c,v}_{\bm{q}}$ with corresponding eigen-energies $\varepsilon^{c,v}_{\bm{q}}$. The Bloch bands are plotted in Fig.~\ref{FigBC}(a). 

The phonons couple to the electronic system through the dependence of the hopping energies $t$ and $t'$ on the lattice displacement $\{\bm{u}\}$. The nearest-neighbor hopping energy $t$ depends on the relative distance $d$ between the two atoms. When the inter-atomic distance changes by $\delta d$ due to atomic displacement, $t$ changes by $\d_d t\delta d$. Here we set $t'$ as a constant for simplicity. 


In this work, we consider the electronic insulating system with the lower Bloch band being completely filled. In the non-interacting case, the many-body ground state $|\Phi_0\rangle$ and the excited one $|\Phi_n\rangle$ can be expressed as the Slater determinant of single-particle states. One thus can calculate the Berry curvature shown in Eq.~\ref{eq6} by using the single-particle states. We can take the Berry curvature induced by the motion of A sublattices along $x$ and $y$ directions as an example, which can be expressed as
\begin{align} \nonumber
&G^{Ax}_{Ay} (\bm{k}) = \frac{i}{N} \sum_{\bm{q}} \frac{\[ {\phi}^{v\dagger}_{\bm{q}}  \mathcal{M}_{\bm{k},Ax} \phi^c_{\bm{q}+\bm{k}}\] \[ {\phi}^{c\dagger}_{\bm{q}+\bm{k}} \mathcal{M}_{\bm{-k},Ay} \phi^v_{\bm{q}}\]} {(\varepsilon^{c}_{\bm{q}+\bm{k}}-\varepsilon^{v}_{\bm{q}})^2} \\ 
\label{eq17} 
& ~~~ - \frac{i}{N} \sum_{\bm{q}} 
\frac{\[ {\phi}^{v\dagger}_{\bm{q}+\bm{k}}  \mathcal{M}_{\bm{-k},Ay} \phi^c_{\bm{q}}\] \[ {\phi}^{c\dagger}_{\bm{q}} \mathcal{M}_{\bm{k},Ax} \phi^v_{\bm{q}+\bm{k}}\]} {(\varepsilon^{c}_{\bm{q}}-\varepsilon^{v}_{\bm{q}+\bm{k}})^2} 
\end{align}
where $\mathcal{M}_{\bm{\pm k},Ax/Ay}$ represent the electron-phonon couplings as detailed in Appendix D that couple the electronic states with a momentum difference of $\bm{k}$.
In Fig. 1(b), we plot the contribution from each electronic momentum $\bm{q}$ to the molecular Berry curvature $G^{Ax}_{Ay} (\bm{k}) $ at $\bm{k}=0$. This corresponds to the phonon induced virtual direct inter-band transition process with the peak contribution concentrating at $K$ and $K'$ valley. 
The dependence of $G^{Ax}_{Ay} (\bm{k}) $ on $\bm{k}$ is plotted in Fig.~\ref{FigBC}(c) where one can find that the Berry curvature shows a peak at the phonon Brillouin zone center. By using reasonably realistic parameters, e.g., $t=3~$eV, $t'=0.02~$eV, $\partial_d t=1~$eV/\AA, we find that the peak of molecular Berry curvature corresponds to an effective magnetic field of $\sim 10^{3}~$Tesla. Thus, the effect of molecular Berry curvature can be large in materials with narrow electronic band gap, e.g., topological materials.
In the small gap limit, we find an analytical formula for the peak value $G^{Ax}_{Ay}(\bm{k}=0)=\frac{3}{2\pi a^2}C(\partial_d t/t)^2$ where $C$ is the Chern number of the system as plotted in the Fig.~\ref{FigBC}(d). 
It is noted that the molecular Berry curvature vanishes at $K$ and $K'$ points due to the vanishing of the matrix elements in $\d_{x,y} \mathcal{H}$ between conduction and valence bands at different valleys in this model.

The formula above can be adopted by the first principle calculation directly, which is thus essential for exploring the phonon Hall effect in magnetic materials.
The $G^{Ax}_{Ay}$ is an analogy of the effective magnetic field in the Raman spin-lattice coupling model~\cite{RamanSpinLattice,Sheng2006, Kagan2008, LZhang, PHE_Heff_Huber_16, PhononModelDiode}. In the Raman spin-lattice coupling model, the motion of $u_{l,Ax}$ can only be influenced by $\dot{u}_{l,Ay}$ on the same site. In contrast, the molecular Berry curvature distributes sharply around the Brillouin zone center. This distribution indicates that, by Fourier transforming back to the real space, the displacement $u_{l,Ax}$ can be influenced by the velocity $\dot{u}_{l',Ay}$ that is far away. 

Moreover, the $G$ matrix also has off-diagonal blocks with nonzero matrix element $G^{Ax}_{By}$, which reflects the correlation between the motions of the A and the B atoms. The amplitude of this term is comparable with $G^{Ax}_{Ay}$ such that $G^{Ax}_{Ay}=-G^{Ax}_{By}$ at $\bm{k}=0$. The off-diagonal block is thus important, which was completely neglected in the traditional Raman spin-lattice coupling, and can lead to a gap opening of the degenerate acoustic bands\cite{TQin2012, Qin2011}. Therefore, the molecular Berry curvature is a better choice to explore the influence of electronic states on phonons in a unified way.

\begin{figure} 
\includegraphics[scale=0.20]{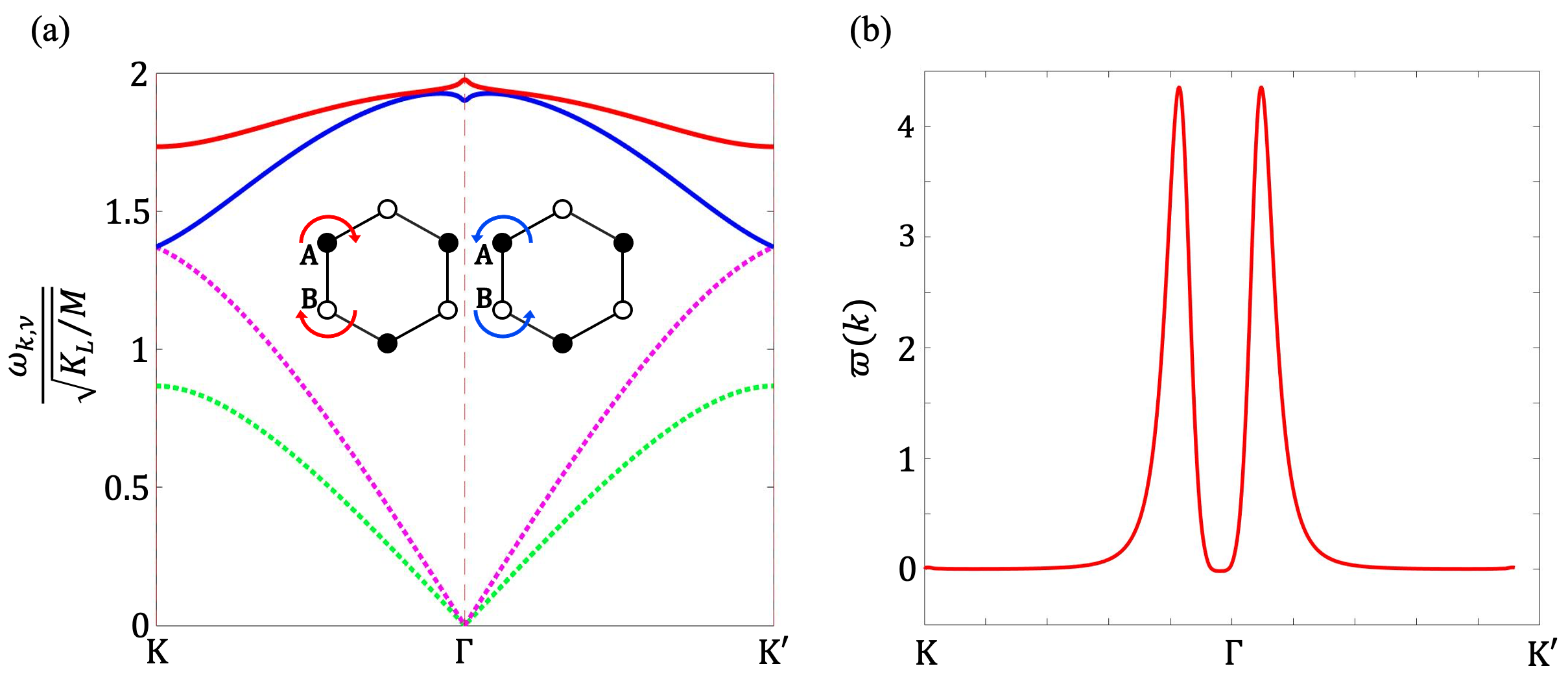}
\caption{(a) Phonon spectrum in the presence of molecular Berry curvature. Gap opening of the optical bands (upper two bands) around the $\Gamma$ point is due to the effect of molecular Berry curvature, where $\delta \omega = \Delta \sqrt{\frac{K_L}{M}}$. Here $K_L = 10^{-3} eV/{\angstrom}^2$ is an in-plane longitudinal and $K_T = K_L/4$ is an in-plane transfers effective spring constants. Inset: Two separate phonon modes corresponding to the frequency at the $\Gamma$ point. The upper phonon band (red color) corresponds to the circular vibrations of atoms in the clockwise direction, and the lower band (blue color) corresponds to the circular vibration in the counter-clockwise direction. The difference of energies of these two modes is $\delta E = \hbar \delta \omega$. (b) Phonon Berry curvature of the upper optical band with the corresponding color plot. The upper and lower optical bands have corresponding Chern numbers of $+1$ and $-1$, respectively. Acoustic bands have zero Chern numbers.} \label{FigBandBC0}
\end{figure}

\subsection{Phonon spectrum and chiral optical phonons}
Once the Berry curvature is calculated for our model we can find the phonon spectrum using Eq. (\ref{eq15}). The result of the numerical calculation of this equation is presented in FIG. 2. In this figure, we can see a gap opening between the optical branches at the $\Gamma$ point. This band gap is proportional to the molecular Berry curvature. To show this relation, we employ the perturbation method since $\tilde{G}^{\dagger}_{\bm{k}} \tilde{G}_{\bm{k}} \ll K_{\bm{k}}$.
From Eq. (\ref{eq14}), one can find that
\begin{align} \label{eq18}
\omega^2_{\bm{k}, \nu} \gamma_{\nu} = K_{\bm{k}} \gamma_{\nu} + 2i\omega_{\bm{k}, \nu} \tilde{G}_{\bm{k}} \gamma_{\nu}.
\end{align}
The second term on the right hand side is treated as a perturbation. The unperturbed eigenvalues $\omega_{1,2}$ and eigenstates $\gamma^0_{1,2}$ for the optical branches at the $\Gamma$ point can be obtained from $\omega_{\bm{k}}^2 \gamma^0_{\nu} = K_{\bm{k}} \gamma^0_{\nu}$. The solutions are $\omega_{1,2} = \omega_{\Gamma}$ and $\gamma^0_{1} = \sqrt{\frac{\omega_0}{4\omega_{\Gamma}}} \(\begin{matrix}1 & 0 & -1 & 0 \end{matrix}\)^T$ and  $\gamma^0_{2} = \sqrt{\frac{\omega_0}{4\omega_{\Gamma}}} \(\begin{matrix} 0 & 1 & 0 & -1 \end{matrix}\)^T$. In the presence of the perturbation, the general eigenstate can be expressed as a linear combination $\tilde{\gamma} = c_1 \gamma^0_1 + c_2 \gamma^0_2$ with $c_1$ and $c_2$ being some constants to be determined. By expanding the phonon eigenvalue at the $\Gamma$ point to the first order as $\omega = \omega_{\Gamma} + \delta \omega$, one can find that
\begin{align} \label{eq19}
\delta \omega \( \begin{matrix} c_1 \\ c_2 \end{matrix} \) = \frac{2i\omega_{\Gamma}}{\omega_0} \( \begin{matrix} {\gamma^0_1}^{\dagger} \tilde{G}_{\bm{k}} \gamma^0_1 & {\gamma^0_1}^{\dagger} \tilde{G}_{\bm{k}} \gamma^0_2 \\ {\gamma^0_2}^{\dagger} \tilde{G}_{\bm{k}} \gamma^0_1 & {\gamma^0_2}^{\dagger} \tilde{G}_{\bm{k}} \gamma^0_2 \end{matrix} \) \( \begin{matrix} c_1 \\ c_2 \end{matrix} \)
\end{align}
where ${\gamma^0_i}^{\dagger} \gamma^0_j = \frac{\omega_0}{2\omega_{\Gamma}} \delta_{ij}$ is employed. Since $\tilde{G}_{\bm{k}}^{\dagger} = - \tilde{G}_{\bm{k}}$, the matrix on the right hand side is Hermitian. Thus the diagonal terms are zero. 
We find that, by setting $c_1 = 1/\sqrt{2}$ and $c_2 = \pm i/\sqrt{2}$, the above matrix can be diagonalized with the phonon energy shifts $\delta \omega = \pm \frac{\hbar}{M}\text{Re} \[G(\bm{k}=0)\]$. Therefore, the optical phonons are split and the phonon polarizations become right- and left-handed polarized. The splitting of the phonon branches at the Brillouin zone center 
is expected to be observable in optical spectral experiments. 

We would like to point out that, in the absence of the molecular Berry curvature, the dynamical matrix can be written as a real matrix at the Brillouin zone \textit{center}. As a result, the phonons are always linearly polarized. In the presence of the molecular Berry curvature, however, phonons at the $\Gamma$ point become circularly polarized. This is different from the chiral phonon at the Brillouin zone \textit{corner}, which has a degenerate state at the opposite momentum.~\cite{Phonon_Chiral_AngularMom_Lifa_15}

By using the phonon wavefunction, one can also define a phonon Berry connection and phonon Berry curvature~\cite{LZhang}. In Fig.~\ref{FigBandBC0}(b), we plot the phonon Berry curvature along the high-symmetric line for the higher optical branch. Peaks appear at the points where the phonon polarization changes from circular around the $\Gamma$ point to linear away from that point. The phonon Berry curvature contributes to a nonzero Chern number $1$. The Chern number for the lower optical branch is $-1$ whereas the acoustic branches have zero Chern numbers. Associated with the spectrum splitting, the Berry curvature of optical branches can also contribute to the thermal Hall effect~\cite{Strohm2005, Inyushkin2007, PHE_SpinLiquid_Exp_17, ChiralPhononCuprate_NP_20}. 


\subsection{Phonon angular momentum}
\begin{figure}
\centering
\includegraphics[scale = 0.25]{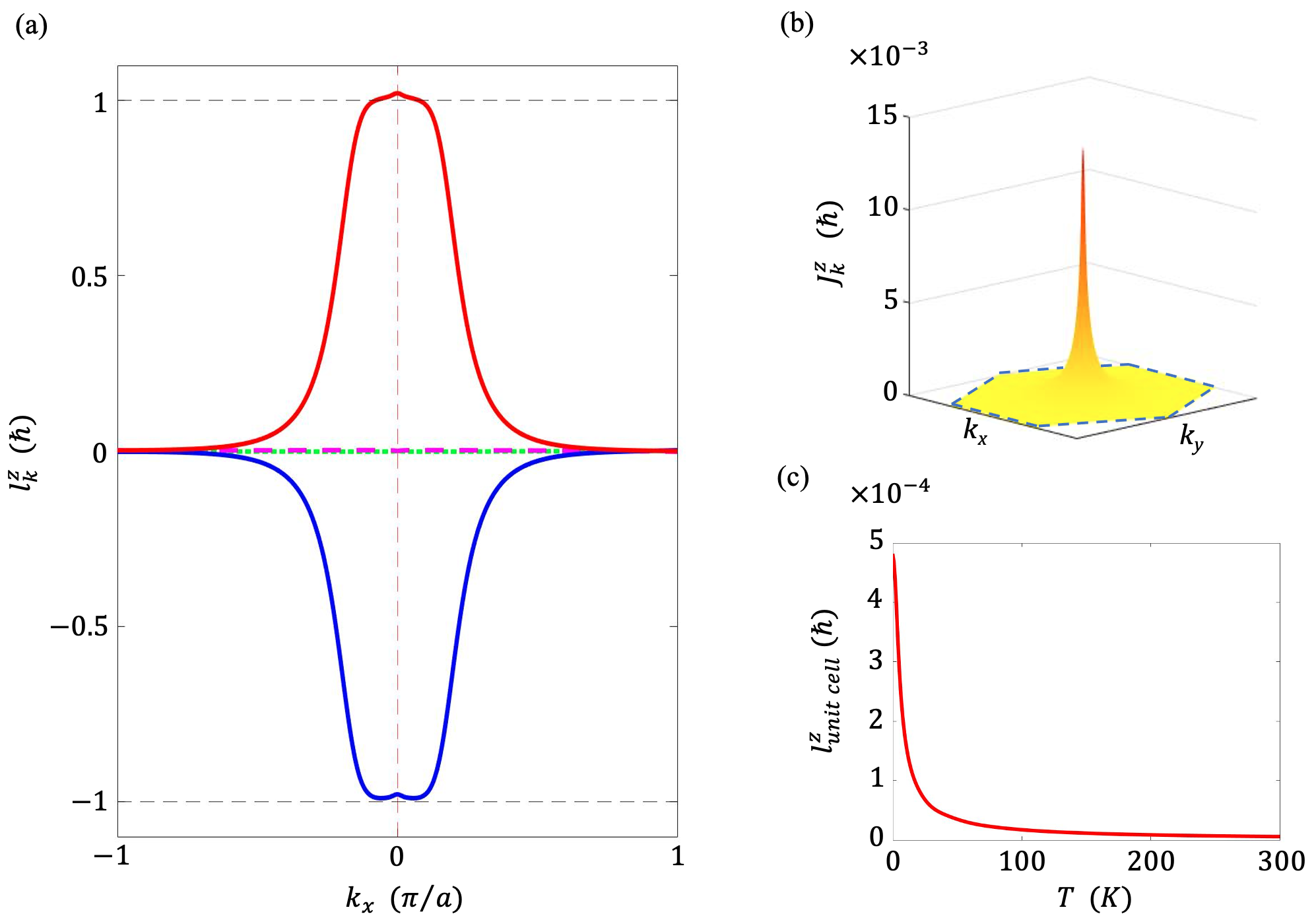}
\caption{(a) Contribution of each phonon band to the total phonon angular momentum, where $l^z_{\bm{k}, \nu}$ represents the angular momentum of each branch without the distribution, such that $\<J_z^\text{ph}\> = \sum_{\bm{k}} J^z(\bm{k}) = \sum_{\bm{k}, \nu} l^z_{\bm{k}, \nu} \Big(\frac{1}{2} + f(\omega_{\bm{k}, \nu})\Big)$ (b) Spectrum of total phonon angular momentum $J^z(\bm{k})$ from all four branches in the limit $T \rightarrow 0 K$. (c) Phonon angular momentum of each unit cell in real space. The angular momentum vanishes in a classical limit $T \rightarrow \infty$.}
\label{Fig3}
\end{figure}

The circular polarization of phonons also give rise to non nonzero phonon angular momentum~\cite{LZhang2014} that can be expressed as
\begin{gather} \label{eq20}
\bm{J}_{\text{ph}} = \sum_{l \alpha} \bm{u}_{l \alpha} \times \bm{\dot{u}}_{l \alpha}.
\end{gather}
For a 2-dimensional system, the vertical component of the angular momentum becomes $J_z^\text{ph} = \sum_{l,\kappa} (u^x_{l\kappa}\dot{u}^y_{l\kappa} - u^y_{l\kappa}\dot{u}^x_{l\kappa})$.  
It can also be written in a matrix product form
\begin{gather} \label{eq21}
J_z^\text{ph} = \sum_{\bm{k}} u^{\dagger}_{\bm{k}} L \dot{u}_{\bm{k}} = \sum_{\bm{k}} u^{\dagger}_{\bm{k}} L \(p_{\bm{k}} + \tilde{G}_{\bm{k}} u_{\bm{k}}\) 
\end{gather}
where $L$ is a real $2r \times 2r$ antisymmetric matrix for a system with $r$ atoms per unit cell. By using the second quantized expression for the canonical variables of atoms (Eqs. (\ref{eq9})-(\ref{eq10})), we calculate the angular momentum for each phonon branch (Fig.~\ref{Fig3}(a)). We find that the phonon angular momentum of both acoustic branches vanish whereas the circularly polarized optical branches are nearly quantized. This is in contrast with the phonon angular momentum obtained by using the Raman spin-lattice model which neglects the \textit{nonlocal} effective magnetic field~\cite{LZhang2014,RamanSpinLatticePhononL}. In those calculations, the acoustic phonons at the Brillouin zone center split and can carry nonzero energy and nonzero angular momentum. The splitting of the acoustics bands is induced by breaking the Galilean translational symmetry due the Raman spin-lattice coupling term that meant to serve as an analogy to a uniformly charged lattice under a real magnetic field. This symmetry is respected by the molecular Berry curvature.

By summing over the angular momentum of all the phonon branches, we find a nonzero value as illustrated in Fig.~\ref{Fig3}(b). This indicates a finite zero-point lattice angular momentum in the zero temperature limit. At finite temperature, the thermal averaged phonon angular momentum is
\begin{align} \label{eq22}
\<J_z^\text{ph}\> = - \sum_{\bm{k}, \nu}  \gamma^{\dagger}_{\nu}L \gamma_{\nu} \( \frac{i2\hbar \omega_{\bm{k}, \nu} }{\omega_0}\) \Big(\frac{1}{2} + f(\omega_{\bm{k}, \nu})\Big) .
\end{align}
Here we used $b_{\bm{k}, \nu} b^{\dagger}_{\bm{k}, \mu} = \delta_{\mu, \nu} + b^{\dagger}_{\bm{k}, \mu} b_{\bm{k}, \nu}$, $\<b^{\dagger}_{\bm{k}, \mu} b^{\dagger}_{\bm{k'}, \nu}\> = \<b_{\bm{k}, \mu} b_{\bm{k'}, \nu}\> = 0$ and $\<b^{\dagger}_{\bm{k}, \mu} b_{\bm{k}, \nu}\> = f(\omega_{\bm{k}, \nu}) \delta_{\mu, \nu}$, where $f(\omega_{\bm{k}, \nu}) = \frac{1}{e^{\beta \hbar \omega_{\bm{k}, \nu}}-1}$ is the Bose-Einstein distribution. In low temperature regime, the phonon angular momentum has a finite value as shown in Fig.~\ref{Fig3}(c). In this limit, an analytic expression for the phonon angular momentum at the $\Gamma$ point can be found by using the perturbative approximation as:
\begin{align} \label{eq23}
    \<J_z^\text{ph}\>_{\Gamma} = \frac{\hbar^2}{M \omega_{\Gamma}}\text{Re}\(G(\bm{k} = 0)\)
\end{align}
which is consistent with the numerical calculations of Eq. (\ref{eq22}). As the temperature increases, the thermal averaged phonon angular momentum tends to go to zero. As $T \rightarrow \infty$, Eq.~\eqref{eq22} approximates to
\begin{align} \label{eq24}
    \<J_z^\text{ph}\> = - \sum_{\bm{k}, \nu}  \gamma^{\dagger}_{\nu}L \gamma_{\nu} \( \frac{2i k_B T }{\omega_0} + \frac{i \hbar^2 \omega^2_{k,\nu}}{6 \omega_0 k_B T}\) 
\end{align}
where the first term vanishes because $\sum_{\nu}  \gamma^{\dagger}_{\nu}L \gamma_{\nu} = 0$ (see Appendix G) and the second term decreases as $1/T$.

\section{SUMMARY} 
We formulated the molecular Berry curvature by using the single-particle Bloch wavefunctions in the absence of a uniform magnetic field. We studied its effect on the lattice dynamics and thus phonons. The quantized equations of motion of the lattice are solved by using the Bogoliubov transformation. We applied our theory to the Haldane model of a honeycomb lattice. For this model, the molecular Berry curvature is narrowly distributed around the Brillouin zone center, which indicates that, in real space, the motion of an ion can be influenced by the velocity of another atom that is far away. This is different from the Lorentz force on the nuclei induced by a magnetic field as well as the widely adopted Raman spin-lattice coupling model. The molecular Berry curvature lifts the degeneracy of optical phonons at the $\Gamma$ point forming chiral phonons with left- and right-handed polarizations. These modes carry nonzero angular momentum and contribute to a nonzero total angular momentum in the low temperature limit and thus modify the Einstein-de Haas effect. These optical branches also carry nonzero phonon Berry curvature that can contribute to the thermal Hall effect.

\section*{ACKNOWLEDGEMENTS}
This work was supported by NSF (EFMA-1641101). We would like thank Junren Shi, Di Xiao and Qiang Gao for helpful discussions.

\newpage

\section*{APPENDIX}

\subsection{Effective lattice Hamiltonian from the time-dependent variational principle} 
The state of electrons is governed by the time-dependent Schr\"{o}dinger equation. By assuming a normalized condition, it can be derived from the time-dependent variational principle with Lagrangian $\mathcal{L}_{\rm{e}}=\langle \Phi_0| i \hbar d_t - H_{\rm{e}} | \Phi_0 \rangle$ by minimizing the action with respect to any variation of $\langle \Phi_0 |$ in the bra space. Under the Born-Oppenheimer approximation, the electronic state lies at the instantaneous ground state of the Hamiltonian $H_e$ that depends on the lattice configuration $\{\bm{R}\}$. With known instantaneous ground state, one can integrate out the electronic degree of freedom to get the effective Lagrangian of lattice that reads
\begin{equation} \label{eq25}
\begin{aligned}
    \mathcal{L} 
    & =\sum_{l,\kappa} \frac{M_{\kappa}}{2}\dot{\bm{R}}_{l,\kappa}^2 + \langle \Phi_0| i\hbar d_t - H_e | \Phi_0 \rangle \\
    & = \sum_{l,\kappa} \frac{M_{\kappa}}{2}\dot{\bm{R}}_{l,\kappa}^2 + \langle \Phi_0| i\hbar d_t | \Phi_0 \rangle  - V_{\rm eff}(\{\bm{R}\}) \\
    & = \sum_{l,\kappa} \frac{M_{\kappa}}{2}\dot{\bm{R}}_{l,\kappa}^2 + \hbar \bm{A}_{l,\kappa}\cdot\dot{\bm{R}}_{l,\kappa} - V_{\rm eff}(\{\bm{R}\}) 
\end{aligned}
\end{equation}
where $\bm{R}_{l,\kappa}$ labels the position of the $\kappa$-th atom in the $l$-th unit cell with mass $M_{\kappa}$. $\bm{A}_{l,\kappa}=\langle \Phi_0| i\nabla_{\bm{R}_{l,\kappa}} | \Phi_0 \rangle$ is the the Berry connection. $V_{\rm eff}(\{\bm{R}\})$ is the total energy of the electrons and ions at the configuration $\{\bm{R}\}$ that forms the potential landscape of the ion. In the equilibrium configuration $\{\bm{R}^0\}$, $V_{\rm eff}$ takes its minimum.
From the Lagrangian, one can reveal the Hamiltonian Eq.~\ref{eq1} by Legendre transformation, which agrees with that derived by Mead and Truhlar~\cite{MeadTruhlar}.

\subsection{The existence of a symmetric gauge near the equilibrium configuration}
We consider a lattice where each atom vibrates around its equilibrium position with a displacement $u_l$ where, in this paragraph, we use shorthand notation for these indices as $ \{ l,\kappa\alpha\} \rightarrow l$ and $\{l', \kappa'\beta\} \rightarrow l'$.
In the small $\{u_l\}$ limit, we can expand the Berry connection $A_{l}=\langle \Phi_0| i\partial_{l} | \Phi_0 \rangle$ to the linear order of $\{u_l\}$ as
$A_l=A_l^0+\partial_{l'} A_l u_{l'}$ where the coefficients $\partial_{l'} A_l$ are taken in the limit of ${\{u_l\}\rightarrow 0}$ and thus are independent of $\{u_l\}$. It is noted that the Berry connection $A_{l}$ is expressed in a parameter space of high dimension. It is questionable whether there exists a gauge transform such that 
$\tilde{A}_l=A_l - \partial_l \chi = -1/2 \sum_{l'} G_{l,l'}u_{l'}$ with gauge invariant $G_{ll'}=\partial_{l}A_{l'}-\partial_{l'}A_{l}=\partial_{l}\tilde{A}_{l'}-\partial_{l'}\tilde{A}_{l}$. The answer is yes. One can first define $\delta A_l=A_l-\tilde{A}_l$. By definition, $\delta A_l=A_l^0+i\sum_{l'}u_{l'}(\frac{1}{2}\langle \partial_{l'}\Phi_0| \partial_{l} \Phi_0 \rangle + \frac{1}{2}\langle \partial_{l}\Phi_0| \partial_{l'} \Phi_0 \rangle+\langle \Phi_0| \partial_{l'} \partial_{l} \Phi_0 \rangle)$. It can be verified that $\partial_l \delta A_{l'} - \partial_{l'} \delta A_{l} = 0$. According to Poincar\'e's Lemma, there always exists locally a scalar function $\chi$ such that $\delta A_l=\partial_l \chi$. One can thus perform such a gauge transformation $e^{i\chi}|\Phi_0 \rangle$ to obtain the Berry connection in the symmetric form.

We can now express the Berry connection (gauge field) in a symmetric gauge. 
The gauge invariant Berry curvature can be written as
\begin{align} \nonumber
&G^{\kappa \alpha}_{\kappa' \beta} (\bm{R}^0_{l}-\bm{R}^0_{l'}) = \[ \frac{\d  A_{l',\kappa'\beta}}{\d u_{l, \kappa\alpha}} - \frac{\d  A_{l,\kappa\alpha}}{\d u_{l', \kappa'\beta}} \]  \\ 
& ~~~~ = i \[\Big\<\frac{\d \Phi_0}{\d u_{l,\kappa\alpha}} \Big| \frac{\d \Phi_0}{\d u_{l', \kappa'\beta}}\Big\> - \Big \< \frac{\d \Phi_0}{\d u_{l',\kappa'\beta}} \Big| \frac{\d \Phi_0}{\d u_{l,\kappa\alpha}} \Big\> \] \label{eq26}
\end{align}
\normalsize
Near the equilibrium position, the Berry connection in the symmetric gauge is
\begin{align} \label{eq27}
A_{\kappa\alpha} (\bm{R}^0_l) = -\frac 12 \sum_{l', \kappa' \beta } G^{\kappa \alpha}_{\kappa' \beta} (\bm{R}^0_{l}-\bm{R}^0_{l'}) u_{\kappa'\beta}(\bm{R}^0_{l'})
\end{align}
and in momentum space:
\begin{align} \label{eq28}
A_{\kappa \alpha} (\bm{k}) = - \sum_{\kappa', \beta} \frac{1}{2} G^{\kappa \alpha}_{\kappa' \beta} (\bm{k}) u_{\kappa'\beta}(\bm{k})
\end{align}

\subsection{Symmetry constraints on the molecular Berry curvature}
In the presence of time reversal symmetry, the Berry curvature $G^{\kappa \alpha}_{\kappa' \beta} (\bm{R}^0_{l}-\bm{R}^0_{l'})=0$ as shown below. We consider the electronic ground state that preserves time reversal invariance and is nondegenerate. Therefore, under time reversal operation $\Theta$, the ground state $| \Phi_0 \rangle$ becomes $| \tilde{\Phi}_e \rangle = |\Theta \Phi_0 \rangle = e^{i\phi}| \Phi_0 \rangle$ with a possible phase difference. Therefore, the Berry connection obtained from $| \tilde{\Phi}_e \rangle$ is $\tilde{A}_{l,\kappa \alpha}=A_{l,\kappa \alpha}+\partial_{l,\kappa \alpha}\phi$. Alternatively, $\tilde{A}_{l,\kappa \alpha}=i\langle \Theta {\Phi}_e| \partial_{l,\kappa \alpha} | \Theta {\Phi}_e \rangle=i(\langle {\Phi}_e| \partial_{l,\kappa \alpha} |  {\Phi}_e \rangle)^*$ by the definition of the time reversal operator $\Theta$ with $^*$ being the complex conjugate. Thus, $\tilde{A}_{l,\kappa \alpha}=-A_{l,\kappa \alpha}$. As a result, $A_{l,\kappa \alpha}=\partial_{l,\kappa \alpha} \phi/2$. The corresponding Berry curvature $G^{\kappa \alpha}_{\kappa' \beta} (\bm{R}^0_{l}-\bm{R}^0_{l'})=\partial_{l,\kappa \alpha} A_{l',\kappa' \beta}-\partial_{l',\kappa' \beta} A_{l,\kappa \alpha}=0$. 

Since the Berry curvature in real space is real number and $G^{\kappa \alpha}_{\kappa' \beta} (\bm{R}^0_{l}-\bm{R}^0_{l'})=-G^{\kappa' \beta}_{\kappa \alpha} (\bm{R}^0_{l'}-\bm{R}^0_{l})$, it can be shown by definition that $G^{\kappa \alpha}_{\kappa' \beta} (\bm{k})=-G^{\kappa' \beta}_{\kappa \alpha} (-\bm{k})=-G^{\kappa' \beta}_{\kappa \alpha} (\bm{k})^*$. Thus, $G(\bm{k})=-G(\bm{k})^\dagger$.

Considering the translational symmetry, one can find that when all the displacement vectors $\bm{u}_{l,\kappa}$ change by the same small amount $\delta \bm{u}$, the Berry connections do not change, i.e., $A_{l,\kappa\alpha}(\{\bm{u}\}+\delta \bm{u})=A_{l,\kappa\alpha}(\{\bm{u}\})$. Thus, $\delta \bm{u} \sum_{l',\kappa'\beta}\partial_{l',\kappa'\beta}A_{l,\kappa\alpha}=0$. Therefore, $\sum_{l',\kappa'\beta}G^{\kappa \alpha}_{\kappa' \beta} (\bm{R}^0_{l}-\bm{R}^0_{l'})=2{\rm{Im}}\sum_{l',\kappa'\beta}\partial_{l',\kappa'\beta}A_{l,\kappa\alpha}=0$. 

\subsection{Molecular Berry curvature in non-interacting electronic system}
The molecular Berry curvature can be expressed in general by the many-body wavefunction $\{\Phi_n\}$ where $n=0$ stands for the ground state and the $n>0$ are excited states 
\begin{widetext}
\begin{equation} \label{eq29}
\begin{aligned}
G^{\kappa \alpha}_{\kappa' \beta} (\bm{k}) &= \frac{1}{N}\sum_{l} \sum_{l'}G^{\kappa \kappa'}_{\alpha \beta} (\bm{R}^0_{l}-\bm{R}^0_{l'}) e^{-i\bm{k}\cdot(\bm{R}^0_{l}-\bm{R}^0_{l'})} \\
&= \frac{1}{N}\sum_{l, l', n\neq0 } i \[\Big\<\frac{\partial \Phi_0}{\partial u_{l, \kappa\alpha}}\Big|\Phi_n\Big\>\Big\<\Phi_n\Big|\frac{\partial \Phi_0}{\partial u_{l', \kappa'\beta}}\Big\>-(u_{l, \kappa\alpha} \leftrightarrow u_{l', \kappa'\beta})\]_{u \rightarrow 0} e^{-i\bm{k}\cdot(\bm{R}^0_{l}-\bm{R}^0_{l'})} \\
& = \frac{i}{N} \sum_{n\neq0} 
\[\frac{\<\Phi_0 |\sum_{l}\frac{\partial H_e}{\partial u_{l, \kappa\alpha}}e^{-i\bm{k}\cdot\bm{R}^0_{l}}|\Phi_n\>\<\Phi_n|\sum_{l'}\frac{\partial H_e}{\partial u_{l', \kappa'\beta}}e^{i\bm{k}\cdot\bm{R}^0_{l'}}|\Phi_0\>}{(E_n-E_0)^2} \] \\
& ~~  ~~~ ~~ ~ - ~ \[\frac{\<\Phi_0 |\sum_{l'}\frac{\partial H_e}{\partial u_{l', \kappa'\beta}}e^{i\bm{k}\cdot\bm{R}^0_{l'}}|\Phi_n\>\<\Phi_n|\sum_{l}\frac{\partial H_e}{\partial u_{l, \kappa\alpha}}e^{-i\bm{k}\cdot\bm{R}^0_{l}}|\Phi_0\>}{(E_n-E_0)^2} \] \\
& = \frac{i}{N}\sum_{n\neq0} 
\frac{\<\Phi_0 |\mathcal{M}_{\bm{k},\kappa\alpha}|\Phi_n\>\<\Phi_n|\mathcal{M}_{-\bm{k},\kappa'\beta}|\Phi_0\> - \<\Phi_0 |\mathcal{M}_{-\bm{k},\kappa'\beta}|\Phi_n\>\<\Phi_n|\mathcal{M}_{\bm{k},\kappa\alpha}|\Phi_0\> }{(E_n-E_0)^2} 
\end{aligned}
\end{equation}
\end{widetext}
where $\mathcal{M}_{\bm{k},\kappa\alpha} = \sum_{l}\frac{\partial H_e}{\partial u_{l, \kappa\alpha}}e^{-i\bm{k}\cdot\bm{R}^0_{l}} =\sqrt{N}\frac{\partial H_{e}}{\partial u_{-\bm{k},\kappa\alpha}} $ in the $u_l\rightarrow 0$ limit. In this limit, the  $\mathcal{M}_{\bm{k},\kappa\alpha}$ involves only single electron scattering process.

In the non-interacting system, the ground state is a product state composed of single-particle states below the chemical potential $\mu$ with $|\Phi_0\rangle=\Pi_{\varepsilon_{m,\bm{q}} < \mu}  c^\dagger_{m,\bm{q}} |0\rangle$ and $c^\dagger_{m,\bm{q}}$ being the creation operator of the state at the momentum $\bm{q}$ of the $m$-th band. The excited states can be expressed $|\Phi_{n'}\rangle$ is the many body states with one occupied excited state and a hole. One can then express the Berry curvature in single particle wavefunction as
\begin{widetext}
\begin{align} \label{eq30}
G^{\kappa \alpha}_{\kappa' \beta} (\bm{k}) = \frac{i}{N} \sum_{\bm{q}}\sum_{\substack{\varepsilon_{m}<\mu \\ \varepsilon_{m'}>\mu}} 
\frac{\phi^\dagger_{m,\bm{q}} \mathcal{M}_{\bm{k},\kappa\alpha}\phi_{m',\bm{q+k}}\phi_{m',\bm{q+k}}^\dagger \mathcal{M}_{-\bm{k},\kappa'\beta}\phi_{m,\bm{q}}}{(\varepsilon_{m,\bm{q}}-\varepsilon_{m',\bm{q+k}})^2} -
\frac{\phi^\dagger_{m,\bm{q+k}}\mathcal{M}_{-\bm{k},\kappa'\beta} \phi_{m',\bm{q}}\phi_{m',\bm{q}}^\dagger \mathcal{M}_{\bm{k},\kappa\alpha}\phi_{m,\bm{q+k}}}{(\varepsilon_{m,\bm{q+k}}-\varepsilon_{m',\bm{q}})^2}
\end{align}
\end{widetext}

\begin{figure}
\centering
\includegraphics[scale = 0.3]{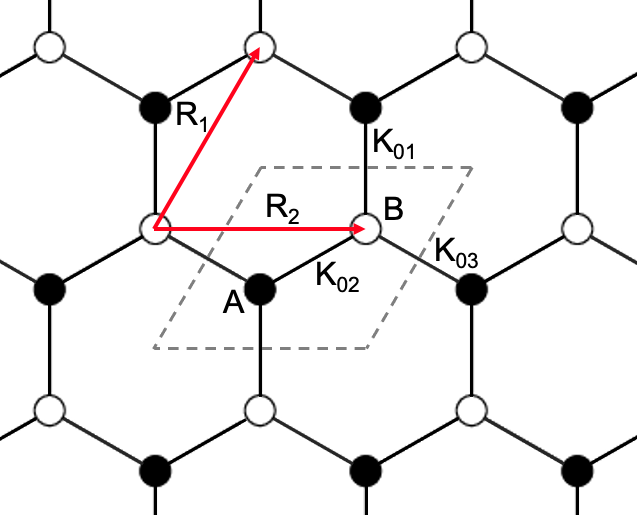}
\caption{Honeycomb model to describe the effective spring constant of a unit cell. In plane longitudinal and transverse spring constants are $K_L$ and $K_T$ respectively}
\label{honeycomb}
\end{figure}

In the following, we focus on the Haldane model to calculate the molecular Berry curvature explicitly. We take $G_{Ay}^{Ax}(\bm{k})$ as an example by setting $u_{l, \kappa\alpha} = u_{l, Ax}$ and $u_{l',\kappa'\beta} = u_{l',Ay}$. For this particular case, we have:
\begin{align*}
\mathcal{M}_{\bm{k},Ax} 
& = \sum_{l} \frac{\d H_e}{\d u_{l, Ax}} e^{-i\bm{k} \cdot \bm{R}^0_{l}} \\
&= \sum_{l, i} 
-(\frac{\d t_i}{\d u_{l, Ax}} b_{l,-R_i}^\dagger a_l + \mathrm{h.c.}) e^{-i\bm{k} \cdot \bm{R}^0_{l}} 
\end{align*}
where $t_i$ with $i=1$-$3$ represents the hopping from site A in the unit cell $\bm{R}_l^0$ to the site B in the unit cell of $\bm{R}_l^0-\bm{R}_i$ with $\bm{R}_1 = (a/2, a\sqrt{3}/2), \bm{R}_2 = (a, 0), \bm{R}_3 = (0, 0)$ as shown in Fig.~\ref{honeycomb}. Here, we have $\frac{\d t_1}{\d u_{l,Ax}} = 0$, $\frac{\d t_2}{\d u_{l, Ax}} = \frac{\sqrt{3}}{2} \d_d t$, $\frac{\d t_3}{\d u_{l,Ax}} = -\frac{\sqrt{3}}{2} \d_d t $ which are independent of $l$ and $\d_d t$ represents the gradient of the hopping energy between two adjacent sites along the bond between them, which we take to be $\d_d t = 1~$eV/{\angstrom}. 

By using the Fourier transformation $a_l=\frac{1}{\sqrt{N}}\sum_{\bm{q}} a_{\bm{q}}e^{i\bm{q}\cdot \bm{R}^0_l}$, we find that 
\begin{align} \nonumber 
\mathcal{M}_{\bm{k},Ax}  & = \frac{\sqrt{3}\d_d t}{2} \sum_{\bm{q}} b_{{\bm{q}}}^\dagger a_{\bm{q+k}} \( e^{i\bm{q} \cdot \bm{R}_2} - e^{-i\bm{q} \cdot \bm{R}_3} \) \\ \nonumber
& + \frac{\sqrt{3}\d_d t}{2} \sum_{\bm{q}} a_{\bm{q}}^\dagger b_{\bm{q+k}} \( e^{-i\bm{(q+k)}\cdot\bm{R}_2} - e^{i\bm{(q+k)} \cdot \bm{R}_3} \) \\
& = \sum_{\bm{q}} \( \begin{matrix}
a^\dagger_{\bm{q}} & b^\dagger_{\bm{q}} \end{matrix} \) \d_x \mathcal{H} \(\begin{matrix}
a_{\bm{q+k}} \\
b_{\bm{q+k}} \end{matrix} \) \label{eq31}
\end{align}
where $\d_x \mathcal{H}$ is a $2\times2$ matrix with only off diagonal elements:
\begin{align*}
& (\d_x \mathcal{H})_{12} =  \frac{\sqrt{3}}{2} \d_d t \( e^{-i\bm{(q+k)}\cdot\bm{R}_2} - e^{i\bm{(q+k)} \cdot \bm{R}_3} \) \\
& (\d_x \mathcal{H})_{21} = \frac{\sqrt{3}}{2} \d_d t \( e^{i\bm{q} \cdot \bm{R}_2} - e^{-i\bm{q} \cdot \bm{R}_3} \).
\end{align*}

Similarly, we have $\frac{\d t_1}{\d u_{l, Ay}} = - \d_d t$, $\frac{\d t_2}{\d u_{l, Ay}} = \frac{\d_d t}{2}$, $\frac{\d t_3}{\d u_{l, Ay}} = \frac{\d_d t}{2}$. 
We thus can obtain that
\begin{align} \nonumber
& \mathcal{M}_{-\bm{k},Ay} = \sum_{l} \frac{\d H_e}{\d u_{l, Ay}} e^{i\bm{k} \cdot \bm{R}^0_{l}} = \\ \nonumber
& = \sum_{l, i} \[ -\frac{\d t_i}{\d u_{l, Ay}} a_{l}^\dagger b_{l+\delta_i} e^{i\bm{k} \cdot \bm{R}^0_{l}} - \frac{\d t_i}{\d u_{l, Ay}} b^\dagger_{l+\delta_i} a_l  e^{i\bm{k} \cdot \bm{R}^0_{l}} \] \\ \nonumber
& =  \frac{\d_d t}{2} \sum_{\bm{q}} a^\dagger_{\bm{k+q}} b_{\bm{q}} \( 2e^{-i \bm{q} \cdot\bm{R}_1} - e^{-i\bm{q}\cdot\bm{R}_2} - e^{i\bm{q} \cdot \bm{R}_3} \) + \\ \nonumber
& + \frac{\d_d t }{2} \sum_{\bm{q}} b_{\bm{k+q}}^\dagger a_{\bm{q}} \( 2e^{i\bm{(k+q)}\cdot\bm{R}_1} - e^{i\bm{(k+q)}\cdot\bm{R}_2} - e^{-i\bm{(k+q)}\cdot\bm{R}_3} \) \\ \label{eq32} 
& = \sum_{\bm{q}} \( \begin{matrix} a^\dagger_{\bm{q+k}} & b^\dagger_{\bm{q+k}} \end{matrix} \) \d_y \mathcal{H} \( \begin{matrix} a_{\bm{q}} \\ b_{\bm{q}} \end{matrix}\)
\end{align}
with:
\begin{align*}
& (\d_y \mathcal{H})_{12} = \frac{\d_d t}{2} \( 2e^{-i \bm{q} \cdot\bm{R}_1} - e^{-i\bm{q}\cdot\bm{R}_2} - e^{i\bm{q} \cdot \bm{R}_3} \) \\
& (\d_y \mathcal{H})_{21} = \frac{\d_d t }{2} \( 2e^{i\bm{(k+q)}\cdot\bm{R}_1} - e^{i\bm{(k+q)}\cdot\bm{R}_2} - e^{-i\bm{(k+q)}\cdot\bm{R}_3} \).
\end{align*}

This is an expression in terms of a single particle wavefunctions and eigenstates. Since only the relative motion of atoms generate the Berry curvature $\bm{u}_{\text{rel}} = \bm{u}_1-\bm{u}_2 \hspace{3mm} \Rightarrow \hspace{3mm} d \bm{u}_1 = -d \bm{u}_2$, where $\bm{u}_1$ and $\bm{u}_2$ are displacements of two different atoms, all 16 different combinations of atoms will generate only 4 independent values of Berry curvature: 
\begin{align*}
 G^{Ax}_{Ax} (\bm{k}) &=  -G^{Ay}_{Ay} (\bm{k}) =  {G^{Bx}_{Bx}}^* (\bm{k}) = -{G^{By}_{By}}^* (\bm{k}) \equiv G_1 (\bm{k})  \\
 {G^{Ax}_{Ay}} (\bm{k}) &= -{G^{Ay}_{Ax}}^* (\bm{k}) = {G^{Bx}_{By}}^* (\bm{k}) = -{G^{By}_{Bx}} (\bm{k}) \equiv G_2(\bm{k})  \\
 {G^{Ax}_{By}} (\bm{k}) &= -{G^{Ay}_{Bx}} (\bm{k}) = {G^{Bx}_{Ay}}^* (\bm{k}) = -{G^{By}_{Ax}}^* (\bm{k}) \equiv G_3(\bm{k})  \\
 G^{Ax}_{Bx} (\bm{k}) &= G^{Ay}_{By} (\bm{k}) = G^{Bx}_{Ax} (\bm{k}) = G^{By}_{Ay} (\bm{k}) \equiv 0
\end{align*}

\subsection{Phonon modes of a honeycomb lattice}

\subsubsection{Hamiltonian for the lattice dynamics}
A semi-classical Hamiltonian of the lattice in the adiabatic approximation can be written in a matrix form as (Eq. \ref{eq1}):
\begin{equation} \label{eq33}
\begin{aligned}  
H_L & = \sum_{l} \frac{1}{2}(p_{l} - \hbar \tilde{A}_{l})^T(p_{l} - \hbar \tilde{A}_{l})+V(\{\bm{u}_l\}) \\
& = \sum_{\bm{k}} \frac{1}{2}(p_{\bm{k}} - \hbar \tilde{A}_{\bm{k}})^{\dagger}(p_{\bm{k}} - \hbar \tilde{A}_{\bm{k}}) + V(\{\bm{u}_{\bm{k}}\}) \\
& = \sum_{\bm{k}} \frac{1}{2}(p_{\bm{k}} + \tilde{G}_{\bm{k}} u_{\bm{k}})^{\dagger} (p_{\bm{k}} + \tilde{G}_{\bm{k}} u_{\bm{k}}) + \sum_{\bm{k}} \frac{1}{2} u_{\bm{k}}^{\dagger} K_{\bm{k}} u_{\bm{k}}
\end{aligned}
\end{equation} 
where the first term is a kinetic energy of atomic vibrations and the latter is the effective interaction of the atoms mediated by the dynamics of electrons. Here the masses have been absorbed into the definition of momentum and displacement vectors. For a lattice with 2 atoms per unit cell, such as a honeycomb lattice, the momentum and displacement vectors can be expressed as:
\begin{align*}
p_{\bm{k}} = \(\begin{matrix} 
\frac{p_{\bm{k},{Ax}}}{\sqrt{M_A}} \\ \frac{p_{\bm{k},{Ay}}}{\sqrt{M_A}} \\ \frac{p_{\bm{k},{Bx}}}{\sqrt{M_B}} \\ \frac{p_{\bm{k},{By}}}{\sqrt{M_B}} \end{matrix}\) ; \hspace{5mm} u_{\bm{k}} =  \(\begin{matrix} \sqrt{M_A} u_{\bm{k},{Ax}} \\ 
\sqrt{M_A} u_{\bm{k},{Ay}} \\ 
\sqrt{M_B} u_{\bm{k},{Bx}} \\ 
\sqrt{M_B} u_{\bm{k},{By}} \end{matrix}\) ;
\end{align*} 
and a gauge field matrix as:
\begin{align*}
\tilde{G}_{\bm{k}} = \frac{\hbar}{2}\( \begin{matrix} \frac{G_1(\bm{k})}{M_A} & \frac{G_2(\bm{k})}{M_A} & 0 & \frac{G_3(\bm{k})}{\sqrt{M_A M_B}} \\ \frac{-G_2(\bm{k})^*}{M_A} & \frac{-G_1(\bm{k})}{M_A} & \frac{-G_3(\bm{k})}{\sqrt{M_A M_B}} & 0 \\ 0 & \frac{G_3(\bm{k})^*}{\sqrt{M_A M_B}} & \frac{G_1(\bm{k})^*}{M_B} & \frac{G_2(\bm{k})^*}{M_B} \\ \frac{-G_3(\bm{k})^*}{\sqrt{M_A M_B}} & 0 & \frac{-G_2(\bm{k})}{M_B} & \frac{-G_1(\bm{k})^*}{M_B} \end{matrix}\)
\end{align*}
which is a skew-Hermitian matrix by definition. Here $K_{\bm{k}}$ is a force constant matrix (in units of $eV/({u \angstrom}^2), u$-atomic mass unit) defined as \cite{LZhang2014}:
\begin{align*} \nonumber
K_{\bm{k}} = \( \begin{matrix}
\frac{K_{01} + K_{02} + K_{03}}{M_A} & -\frac{K_{02} + K_{01}e^{-i\bm{k \cdot R_1}} + K_{03}e^{-i\bm{k \cdot R_2}}}{\sqrt{M_A M_B}} \\
-\frac{K_{02} + K_{01}e^{i\bm{k \cdot R_1}} + K_{03}e^{i\bm{k \cdot R_2}}}{\sqrt{M_A M_B}} & \frac{K_{01} + K_{02} + K_{03}}{M_B}
\end{matrix} \)
\end{align*}
where $\bm{k \cdot R_1} = k_x a/2 + \sqrt{3}k_y a/2$ and $\bm{k \cdot R_2} = k_x a$, with $a$ being a distance between two neighboring unit cells with unit vectors $(a, 0)$ and $(a/2, a \sqrt{3}/2)$. Here $K_{01} = U(\pi/2) K_x U(-\pi/2)$, $K_{02} = U(\pi/6) K_x U(-\pi/6)$ and $K_{03} = U(-\pi/6) K_x U(\pi/6)$ where $K_x = \( \begin{matrix} K_L & 0 \\ 0 & K_T \end{matrix} \)$ is a spring constant matrix constructed from longitudinal and transverse spring constants $K_L$ and $K_T$, and $U(\theta) = \( \begin{matrix} \cos{\theta} & -\sin{\theta} \\ \sin{\theta} & \cos{\theta}\end{matrix} \)$ is a 2-dimensional rotation operator in $x-y$ plane. Combining all these we obtain the lattice Hamiltonian as in Eq. (\ref{eq33}).

Considering all these the lattice Hamiltonian can be written as:

\begin{align} \label{eq34}
H_L = \sum_{\bm{k}} \frac 12 \[ p_{\bm{k}}^{\dagger} p_{\bm{k}} +  u^{\dagger}_{\bm{k}}D_{\bm{k}} u_{\bm{k}} + (p_{\bm{k}}^{\dagger} \tilde{G}_{\bm{k}} u_{\bm{k}} + h.c.) \]
\end{align}

where $D_{\bm{k}} = K_{\bm{k}} + \tilde{G}^{\dagger}_{\bm{k}} \tilde{G}_{\bm{k}}$. We then can get a pair of canonical equations of motion
\begin{align} \label{eq35}
&\dot{p}_{\bm{k}} = - \frac{\d H}{\d u_{-\bm{k}}} = \tilde{G}_{\bm{k}} p_{\bm{k}} - D_{\bm{k}} u_{\bm{k}} \\ \label{eq36}
&\dot{u}_{\bm{k}} = ~~ \frac{\d H}{\d {p}_{-\bm{k}}} = p_{\bm{k}} + \tilde{G}_{\bm{k}} u_{\bm{k}} .
\end{align}

\subsubsection{Second quantization with Bogoliubov transformation}
After introducing the second quantization of displacement and momentum as \cite{Kittel1987} $u_{\bm{k}} = \sqrt{\frac{\hbar}{2 \omega_0}} \(a^{\dagger}_{-\bm{k}} + a_{\bm{k}}\) = \sqrt{\frac{\hbar}{2 \omega_0}} \bar{u}_{\bm{k}}$ and $p_{\bm{k}} = i \sqrt{\frac{\hbar \omega_0 }{2}} \(a^{\dagger}_{-\bm{k}} - a_{\bm{k}}\) = \sqrt{\frac{\hbar \omega_0 }{2}} \bar{p}_{\bm{k}}$, where $a^{\dagger}_{-\bm{k}}$ and $a_{\bm{k}}$ represent column vectors of creation and annihilation operators, the canonical equations of motion can be combined into a matrix form:
\begin{align} \label{eq37}
\( \begin{matrix} 
\dot{\bar{u}}_{\bm{k}} \\
\dot{\bar{p}}_{\bm{k}} \end{matrix} \) = \(\begin{matrix} 
\tilde{G}_{\bm{k}} & \mathbb{1} \omega_0 \\
-\frac{D_{\bm{k}}}{\omega_0} & \tilde{G}_{\bm{k}} \end{matrix} \) \( \begin{matrix}
\bar{u}_{\bm{k}} \\
\bar{p}_{\bm{k}} \end{matrix} \) 
\end{align}
Now replacing $\( \begin{matrix} 
\bar{u}_{\bm{k}} \\ 
\bar{p}_{\bm{k}}
\end{matrix} \) = \( \begin{matrix}
\mathbb{1} & \mathbb{1} \\
\mathbb{1}i & -\mathbb{1}i 
\end{matrix} \) \( \begin{matrix} 
a^{\dagger}_{-\bm{k}} \\ 
a_{\bm{k}}
\end{matrix} \)$ we can obtain:
\begin{widetext}
\begin{align} \label{eq38}
\( \begin{matrix}
-\mathbb{1}i & 0 \\
0 & \mathbb{1}i 
\end{matrix} \) \( \begin{matrix} 
\dot{\tilde{a}}^{\dagger}_{-\bm{k}} \\ 
\dot{a}_{\bm{k}}
\end{matrix} \) = \frac 12 \( \begin{matrix} 
\frac{D_{\bm{k}}}{\omega_0} + \mathbb{1} \omega_0 - 2 i \tilde{G}_{\bm{k}} & \frac{D_{\bm{k}}}{\omega_0} - \mathbb{1} \omega_0 \\ 
\frac{D_{\bm{k}}}{\omega_0} - \mathbb{1} \omega_0 & \frac{D_{\bm{k}}}{\omega_0} + \mathbb{1} \omega_0 + 2 i \tilde{G}_{\bm{k}} 
\end{matrix} \) \( \begin{matrix} 
\tilde{a}^{\dagger}_{-\bm{k}} \\ 
a_{\bm{k}}
\end{matrix} \) = \tilde{\Omega}^*_{\bm{k}} \( \begin{matrix}
\tilde{a}^{\dagger}_{-\bm{k}} \\ 
a_{\bm{k}}
\end{matrix} \)
\end{align}
\end{widetext}
where $\tilde{\Omega}^*_{\bm{k}}$ is an $8\times8$ positive semi-definite Hermitian matrix. The notations $\tilde{\Omega}^*_{\bm{k}}$ was chosen for the convenience that will be clear shortly. Now we introduce the Bogoliubov transformation as \cite{FetterWalecka1971}:
\begin{align} \label{eq39}
&a_{\bm{k}} = \sum_{\nu} \( \alpha_{\nu} b_{\bm{k}, \nu} + \beta^*_{\nu} b^{\dagger}_{-\bm{k}, \nu} \) \\ \label{eq40}
&\tilde{a}^{\dagger}_{-\bm{k}} = \sum_{\nu} \( \alpha^*_{\nu} b^{\dagger}_{-\bm{k}, \nu} + \beta_{\nu} b_{\bm{k}, \nu} \)
\end{align} 
where the tilde represents the transpose of the vector and the summation is over all the branches. Here $b_{\bm{k}, \nu}$ $(b^{\dagger}_{-\bm{k}, \nu})$ are single valued Bogoliubov operators corresponding to each branch and $\alpha_{\nu}$ and $\beta_{\nu}$ are column vectors of 4 elements corresponding to each degree of freedom. We require that each Bogoliubov operators represent the eigenstates with a specific frequency $\omega_{\bm{k}, \nu}$, such that $\dot{b}_{\bm{k}, \nu} = -i\omega_{\bm{k}, \nu} b_{\bm{k}, \nu}$ and $\dot{b}^{\dagger}_{-\bm{k}, \nu} = i\omega_{-\bm{k}, \nu} b^{\dagger}_{-\bm{k}, \nu}$. Using this transformation we can obtain the following equations:
\begin{align} \label{eq41}
& \omega_{\bm{k}, \nu} \sigma_z \chi = \tilde{\Omega}_{\bm{k}} \chi \\ \label{eq42}
& \omega_{-\bm{k}, \nu} \sigma_z \chi^* = \tilde{\Omega}^*_{-\bm{k}} \chi^*
\end{align}
where $\chi = \( \begin{matrix} \alpha_{\nu} \\ \beta_{\nu}  \end{matrix} \)$ and $\tilde{\Omega}_{\bm{k}}$ is same as was defined in Eq. (\ref{eq41}). Now, if we introduce a new eigenstate as $\tilde{\chi} = \tilde{\Omega}_{\bm{k}}^{1/2} \chi$, we can obtain a new eigenvalue equation as
\begin{align} \label{eq43}
\omega_{\bm{k, \nu}} \tilde{\chi} = \tilde{\Omega}^{1/2}_{\bm{k}} \sigma_z \tilde{\Omega}^{1/2}_{\bm{k}} \tilde{\chi} = \tilde{H}_{\text{eff}} \tilde{\chi}
\end{align}
where $\tilde{H}_{\text{eff}}$ is an effective Hamiltonian which is an $8\times8$ Hermitian matrix and can be solved to find the $\omega_{\bm{k}, \nu}$.

As Eq. (\ref{eq43}) suggests, we will obtain 8 different phonon branches but only 4 of them should have physical meaning as we have only 4 physical degrees of freedom. In the following discussion we will show how to pick those 4 physical branches. Using Eq. (\ref{eq41}) we can get a relation:
\begin{align} \label{eq44}
\omega_{\bm{k}, \nu} \(\alpha^{\dagger}_{\nu} \alpha_{\nu} - \beta^{\dagger}_{\nu} \beta_{\nu}\) = \( \begin{matrix} \alpha^{\dagger}_{\nu} & \beta^{\dagger}_{\nu} \end{matrix} \) \tilde{\Omega}_{\bm{k}} \( \begin{matrix}
\alpha_{\nu} \\
\beta_{\nu} 
\end{matrix} \)
\end{align}
This expression will help us to identify the 4 branches we are looking for. For that, we first need to put constraints on the Bogoliubov transformation. Initial bosonic operators had the commutation relations as
\begin{align} \label{eq45}
& \[a_{\bm{k}, d}, a^{\dagger}_{\bm{k'}, d'}\] = \delta_{\bm{k k'}} \delta_{d d'} \\  \label{eq46}
& \[a_{\bm{k}, d}, a_{\bm{k'}, d'}\] = \[a^{\dagger}_{\bm{k}, d}, a^{\dagger}_{\bm{k'}, d'}\] = 0
\end{align} 
After the transformation we require that the new operators should obey similar bosonic commutation relations. For that we write
\begin{align} \label{eq47}
& \[b_{\bm{k}, \nu}, b^{\dagger}_{\bm{k'}, \mu}\] = \delta_{\bm{k k'}} \delta_{\mu \nu} \\ \label{eq48}
& \[b_{\bm{k}, \nu}, b_{\bm{k'}, \mu}\] = \[b_{\bm{k}, \nu}^{\dagger}, b_{\bm{k'}, \mu}^{\dagger}\] = 0
\end{align}
From these two conditions it is easy to show the following relations:
\begin{align} \label{eq49}
& \sum_{\nu} \( \alpha^{\nu}_{d} {\alpha^{\nu}_{d'}}^* - {\beta^{\nu}_{d}}^* \beta^{\nu}_{d'}\) = \delta_{d d'} \\ \label{eq50}
& \sum_d \( {\alpha^{\mu}_{d}}^* \alpha^{\nu}_{d} - {\beta^{\nu}_{d}}^* \beta^{\mu}_{d}\) = \delta_{\mu \nu}
\end{align}
where the first one can be defined as the completeness relation and the second one as orthonormal condition. From this we can see that for the Bogoliubov transformations to preserve the bosonic commutation relations we should have $\alpha^{\dagger}_{\nu} \alpha_{\nu} - \beta^{\dagger}_{\nu} \beta_{\nu} = +1$, i.e it should be a positive number, and this appears on the left hand side of Eq (\ref{eq50}). It can also be shown that $\Omega_{\bm{k}}$ is a positive semidefinite matrix and we conclude that only positive solutions of $\omega_{\bm{k}, \nu}$ should be considered as physical.\\

Alternatively, if we switch to a new basis as $\gamma_{\nu} = \frac{1}{\sqrt{2}} \(\alpha_{\nu} + \beta_{\nu} \)$ and $\bar{\gamma}_{\nu} = \frac{1}{\sqrt{2}} \(\alpha_{\nu} - \beta_{\nu}\)$ Eq. (\ref{eq44}) can be rewritten as
\begin{align} \label{eq51}
\omega_{\bm{k}, \nu} \(\begin{matrix} \gamma_{\nu} \\ \bar{\gamma}_{\nu} \end{matrix}\) = \(\begin{matrix} i \tilde{G}_{\bm{k}} & \omega_0 \\ \frac{D_{\bm{k}}}{\omega_0} & -i \tilde{G}^{\dagger}_{\bm{k}} \end{matrix}\) \(\begin{matrix} \gamma_{\nu} \\ \bar{\gamma}_{\nu} \end{matrix}\)
\end{align}
with the normalization condition resulted from Eq. (\ref{eq44}):
\begin{align} \label{eq52}
\gamma^{\dagger}_{\nu} \bar{\gamma}_{\nu} + \bar{\gamma}^{\dagger}_{\nu} \gamma_{\nu} = 1
\end{align}
From this we can construct a more compact eigenvalue problem: multiplying both sides of Eq. (\ref{eq57}) by $\sigma_x$ we obtain
\begin{align} \label{eq53}
\omega_{\bm{k}, \nu} \sigma_x \psi_{\nu} = \(\begin{matrix} \frac{D_{\bm{k}}}{\omega_0} & -i \tilde{G}^{\dagger}_{\bm{k}} \\ i \tilde{G}_{\bm{k}} & \omega_0 \end{matrix}\) \psi_{\nu} = \Omega_{\bm{k}} \psi_{\nu}
\end{align}
where $\psi_{\nu} = \(\begin{matrix} \gamma_{\nu} \\ \bar{\gamma}_{\nu}\end{matrix}\)$. Here $\Omega_{\bm{k}}$ is again a semi-definite positive matrix and for that we can introduce new eigenstate as $\tilde{\psi}_{\nu} = \Omega_{\bm{k}}^{1/2} \psi_{\nu}$ and obtain:
\begin{align} \label{eq54}
\omega_{\bm{k}, \nu} \tilde{\psi}_{\nu} = \Omega_{\bm{k}}^{1/2} \sigma_x \Omega_{\bm{k}}^{1/2} \tilde{\psi}_{\nu} = H_{\text{eff}} \tilde{\psi}_{\nu} 
\end{align}
where effective Hamiltonian $H_{\text{eff}}$ introduced above is Hermitian and can be solved to find the eigenvalues $\omega_{\bm{k}, \nu}$. 
The result of numerical calculation of this equation is same as the one obtained from Eq. (\ref{eq46}). 

\subsection{Determinant of the eigenvalues and related}\label{determinant}
The effective Hamiltonian can be written as $H_{\rm eff}=\Omega^{1/2}_{\bm{k}}(\sigma_x\otimes I_d)\Omega^{1/2}_{\bm{k}}$ where $I_d$ is the identity matrix with the dimension $d$ being that of the $K$ matrix. Thus the determinant $\det H_{\rm eff} = \det \Omega \cdot (\det \sigma_x)^d$. From the definition of the $\Omega$, we can find that $\det \Omega = \det (\omega_0 I_d) \cdot \det(D/\omega_0 - iG(\omega_0 I_d)^{-1}iG)=\det (D+G^2)=\det(D-GG^\dagger)=\det K$.

By the definition of $\Omega$, one can find that $\Omega_{-\bm{k}}=U^\dagger \Omega_{\bm{k}}^* U$ with $U=\sigma_z$. Thus, $H_{\rm eff}(-\bm{k})=(U^\dagger \Omega_{\bm{k}}^* U)^{1/2}\sigma_x(U^\dagger \Omega_{\bm{k}}^* U)^{1/2}$. By noting that $(U^\dagger \Omega_{\bm{k}}^* U)^{1/2}=U^\dagger \Omega_{\bm{k}}^{*1/2} U$, one can find that $H_{\rm eff}(-\bm{k})=-U^\dagger H_{\rm eff}(\bm{k})^* U$.

\subsection{Phonon angular momentum}
A classical angular momentum phonons is defined as:
\begin{align} \nonumber
J_z^\text{ph} &= \sum_{l,\kappa} (u^x_{l\kappa}\dot{u}^y_{l\kappa} - u^y_{l\kappa}\dot{u}^x_{l\kappa}) \\ \nonumber
&= \sum_{l,\kappa}  \( \begin{matrix}
u_{l\kappa}^x \\
u_{l\kappa}^y 
\end{matrix} \)^{T} \( \begin{matrix}
0   & 1 \\
-1   &  0 
\end{matrix} \) \( \begin{matrix}
\dot{u}_{l\kappa}^x \\
\dot{u}_{l\kappa}^y 
\end{matrix} \) \\ \label{eq55}
& = \sum_{\bm{k},\kappa} \( \begin{matrix}
u^{\kappa, x}_{\bm{k}} \\
u^{\kappa, y}_{\bm{k}} 
\end{matrix} \)^{\dagger} \( \begin{matrix}
0   & 1 \\
-1   &  0 
\end{matrix} \) \( \begin{matrix}
\dot{u}^{\kappa, x}_{\bm{k}} \\
\dot{u}^{\kappa, y}_{\bm{k}} 
\end{matrix} \)
\end{align}
For a system with n = 2 atoms per unit cell, such as honeycomb lattice, the total phonon angular momentum can be written as:
\begin{align} \nonumber
J_z^\text{ph} &= \sum_{\bm{k}} 
\( \begin{matrix}
u^{A, x}_{\bm{k}} \\
u^{A, y}_{\bm{k}} \\
u^{B, x}_{\bm{k}} \\
u^{B, y}_{\bm{k}}
\end{matrix} \)^{\dagger} \( \begin{matrix}
0 & 1 & 0 & 0 \\
-1 & 0 & 0 & 0\\
0 & 0 & 0 & 1\\
0 & 0 & -1 & 0
\end{matrix} \) \( \begin{matrix}
\dot{u}^{A, x}_{\bm{k}}  \\
\dot{u}^{A, y}_{\bm{k}}  \\
\dot{u}^{B, x}_{\bm{k}}  \\
\dot{u}^{B, y}_{\bm{k}}
\end{matrix} \) \\ \label{eq56}
&= \sum_{\bm{k}} u^{\dagger}_{\bm{k}} L \dot{u}_{\bm{k}} = \sum_{\bm{k}} u^{\dagger}_{\bm{k}} L \(p_{\bm{k}} + G_{\bm{k}} u_{\bm{k}}\) 
\end{align}
We replace the canonical variables with $u_{\bm{k}} = \sum_{\nu} \sqrt{\frac{\hbar}{\omega_0}} \( \gamma^*_{\nu} b^{\dagger}_{-\bm{k}, \nu} + \gamma_{\nu} b_{\bm{k}, \nu} \)$ and $p_{\bm{k}} = \sum_{\nu} i \sqrt{\hbar \omega_0} \( \bar{\gamma}^*_{\nu} b^{\dagger}_{-\bm{k}, \nu} - \bar{\gamma}_{\nu} b_{\bm{k}, \nu}\)$ using which we can get the expression for the phonon angular momentum as:
\begin{align} \nonumber
J_z^\text{ph} &= \sum_{\bm{k}, \mu, \nu} \hbar \( i\gamma^{T}_{\mu}L\bar{\gamma}^{*}_{\nu} + \frac{1}{\omega_0} \gamma^{T}_{\mu}L G_{\bm{k}}\gamma^{*}_{\nu} \) b_{-\bm{k}, \mu} b^{\dagger}_{-\bm{k}, \nu} \\ \nonumber
&+ \sum_{\bm{k}, \mu, \nu} \hbar \( -i \gamma^{\dagger}_{\mu}L\bar{\gamma}_{\nu} + \frac{1}{\omega_0} \gamma^{\dagger}_{\mu} L G_{\bm{k}} \gamma_{\nu} \)b^{\dagger}_{\bm{k}, \mu} b_{\bm{k}, \nu} \\ \nonumber
&+ \sum_{\bm{k}, \mu, \nu} \hbar \( -i \gamma^{T}_{\mu} L \bar{\gamma}_{\nu} + \frac{1}{\omega_0} \gamma^{T}_{\mu} L G_{\bm{k}} \gamma_{\nu} \)b_{-\bm{k}, \mu} b_{\bm{k}, \nu} \\ \label{eq57}
&+ \sum_{\bm{k}, \mu, \nu} \hbar \( i\gamma^{\dagger}_{\mu} L \bar{\gamma}^{*}_{\nu} + \frac{1}{\omega_0} \gamma^{\dagger}_{\mu}L G_{\bm{k}} \gamma^{*}_{\nu} \) b^{\dagger}_{\bm{k}, \mu} b^{\dagger}_{-\bm{k}, \nu}
\end{align}
We can calculate the thermal average of this expression. Since $b_{\bm{k}, \nu} b^{\dagger}_{\bm{k}, \mu} = \delta_{\mu, \nu} + b^{\dagger}_{\bm{k}, \mu} b_{\bm{k}, \nu} $ and $\<b^{\dagger}_{\bm{k}, \mu} b_{\bm{k}, \nu}\> = f(\omega_{\bm{k}, \nu}) \delta_{\mu, \nu}$, $\<b^{\dagger}_{\bm{k}, \mu} b^{\dagger}_{\bm{k'}, \nu}\> = \<b_{\bm{k}, \mu} b_{\bm{k'}, \nu}\> = 0$, where $f(\omega_{\bm{k}, \nu}) = \frac{1}{e^{\beta \hbar \omega_{\bm{k}, \nu}}-1}$ is the Bose-Einstein distribution, we can write: 
\begin{align} \nonumber
\<J_z^\text{ph}\> &= \sum_{\bm{k}, \nu} \hbar \( i \gamma^{T}_{\nu} L \bar{\gamma}^{*}_{\nu} + \frac{1}{\omega_0} \gamma^{T}_{\nu} L G_{\bm{k}}\gamma^{*}_{\nu} \) \(1 + f(\omega_{\bm{k}, \nu})\) \\ \nonumber
&+ \sum_{\bm{k}, \nu} \hbar \( -i \gamma^{\dagger}_{\nu}L\bar{\gamma}_{\nu} + \frac{1}{\omega_0} \gamma^{\dagger}_{\nu} L G_{\bm{k}}\gamma_{\nu} \) f(\omega_{\bm{k}, \nu}) \\ \nonumber
&= \sum_{\bm{k}, \nu} \hbar \( \frac{1}{\omega_0} \gamma^{\dagger}_{\nu} L G_{\bm{k}}\gamma_{\nu} -i \gamma^{\dagger}_{\nu}L\bar{\gamma}_{\nu} \) \(1 + 2 f(\omega_{\bm{k}, \nu})\) \\ \nonumber
&= \sum_{\bm{k}, \nu} \hbar \gamma^{\dagger}_{\nu}L \( \frac{\tilde{G}_{\bm{k}}}{\omega_0}\gamma_{\nu} - i\bar{\gamma}_{\nu}\) \(1 + 2 f(\omega_{\bm{k}, \nu})\) \\ \label{eq58}
&= - \sum_{\bm{k}, \nu}  \gamma^{\dagger}_{\nu}L \gamma_{\nu} \( \frac{i\hbar \omega_{\bm{k}, \nu} }{\omega_0}\) \(1 + 2 f(\omega_{\bm{k}, \nu})\) 
\end{align}

Here we show that $\sum_\nu \gamma^{\dagger}_{\nu}L \gamma_{\nu} =0$. 
\begin{align}
\sum_\nu \gamma^{\dagger}_{\nu}L \gamma_{\nu} 
& =  {\rm Tr} {\psi}^\dagger \hat{L} {\psi} \nonumber \\
& = {\rm Tr} \tilde{\psi}^\dagger  \tilde{\Omega}_{\bm{k}}^{-1/2} \hat{L} \omega_{\bm{k}, \nu} \tilde{\Omega}_{\bm{k}}^{-1/2} \tilde{\psi} \nonumber \\
& =  {\rm Tr} \tilde{\psi}^\dagger  \Omega_{\bm{k}}^{-1/2} \hat{L} \sigma_x \Omega_{\bm{k}}^{1/2} \tilde{\psi} \nonumber \\
& = {\rm Tr} \Omega_{\bm{k}}^{1/2} \tilde{\psi} \tilde{\psi}^\dagger  \Omega_{\bm{k}}^{-1/2} \hat{L} \sigma_x \nonumber \\ \label{eq59}
& = {\rm Tr}  \hat{L} \sigma_x = 0
\end{align}
where $\omega_{\bm{k}, \nu}$ is the eigen-energy, which is a diagonal matrix, and $\hat{L} =  \( \begin{matrix}
              L & 0 \\ 
              0 & 0 
            \end{matrix} \) $

\end{document}